\documentclass[epj,nopacs]{svjour}
\usepackage[varg]{txfonts}

\usepackage{amsmath,amssymb}
\usepackage{amssymb}
\usepackage{amsfonts}
\usepackage[normalem]{ulem}
\usepackage{textcomp}
\usepackage{hyperref}
\usepackage{enumitem}
\usepackage{bm}
\usepackage{afterpage}
\usepackage{graphicx}
\usepackage{psfrag}
\usepackage{mathtools}
\usepackage{layouts}
\usepackage{epstopdf}
\usepackage[usenames,dvipsnames]{xcolor}
\usepackage[utf8]{inputenc}
\usepackage{multirow}
\usepackage{rotating}
\usepackage{tabularx}
\usepackage{ragged2e}
\usepackage{blindtext}
\usepackage{graphicx}
\usepackage{siunitx}
\usepackage{float}

\usepackage{cite}

\begin{document}

	
	\title{Maximum Entropy Spectral Analysis: an application to gravitational waves data analysis}
	\author{Alessandro~{Martini}\inst{1,2,3}\thanks{alessandro.martini-1@unitn.it} \and 
	Stefano~{Schmidt}\inst{4,5}\thanks{s.schmidt@uu.nl} \and Gregory~{Ashton}\inst{6} \thanks{Gregory.Ashton@rhul.ac.uk}\and Walter~{Del Pozzo}\inst{1}\thanks{walter.delpozzo@unipi.it}}
        
        \institute{Dipartimento di Fisica  Università di Pisa, and INFN Sezione di Pisa, Pisa I-56127,Italy
        		      \and Università di Trento, Dipartimento di Fisica, I-38123 Povo, Trento, Italy
		      \and INFN, Trento Institute for Fundamental Physics and Applications, I-38123 Povo, Trento, Italy
        		      \and Institute for Gravitational and Subatomic Physics (GRASP),Utrecht University, Princetonplein 1, 3584 CC, Utrecht, The Netherlands
		      \and Nikhef, Science Park 105, 1098 XG, Amsterdam, The Netherlands
		      \and Royal Holloway University of London, London, United Kingdom}

\abstract{
The Maximum Entropy Spectral Analysis (MESA) method, developed by Burg, offers a powerful tool for spectral estimation of a time-series. It relies on Jaynes’ maximum entropy principle, allowing the spectrum of a stochastic process to be inferred using the coefficients of an autoregressive process AR($p$) of order $p$. A closed-form recursive solution provides estimates for both the autoregressive coefficients and the order $p$ of the process. We provide a ready-to-use implementation of this algorithm in a Python package called \texttt{memspectrum}, characterized through power spectral density (PSD) analysis on synthetic data with known PSD and comparisons of different criteria for stopping the recursion. Additionally, we compare the performance of our implementation with the ubiquitous Welch algorithm, using synthetic data generated from the GW150914 strain spectrum released by the LIGO-Virgo-Kagra collaboration. Our findings indicate that Burg’s method provides PSD estimates with systematically lower variance and bias. This is particularly manifest in the case of a small (O($5000$)) number of data points, making Burg’s method most suitable to work in this regime. Since this is close to the typical length of analysed gravitational waves data, improving
the estimate of the PSD in this regime leads to more reliable posterior profiles for the system under study. We conclude
our investigation by utilising MESA, and its particularly easy parametrisation where the only free parameter is the order
p of the AR process, to marginalise over the interferometers noise PSD in conjunction with inferring the parameters of
GW150914 
}

\titlerunning{MESA: an application to gravitational waves data analysis}
\authorrunning{Martini, Schmidt, Ashton, Del Pozzo}
	
	\maketitle


\section{Introduction}

The problem of inferring the morphology and the defining parameters of deterministic signals superimposed to stochastic processes is one of the most wide spread and interesting problems in several areas of human activities. Whenever some form of model for the signal we are looking for is available, the problem is typically solved via the Wiener filter, defined as the whitening filter that maximises the signal-to-noise ratio, i.e. the relative power of the (known) signal over the power of the (known) underlying stochastic process. Hence, signal detection and characterisation requires accurate knowledge of i) the shape of the signal we are looking for and ii) the statistical properties of the stochastic process. The construction of signal models is typically driven either by physical or by mathematical arguments hence, although extremely difficult in general, it is doable. On the other hand, stochastic process models can be extremely difficult to construct, both for practical and theoretical reasons. A stochastic process is fully described by the knowledge of the probability distribution governing its realisations -- the ``paths'' of the random variable under scrutiny -- over the entire time axis, from $t=-\infty$ to $t=\infty$. Clearly this is not possible in practice. Therefore modelling a stochastic process either relies on modelling of the underlying physical processes, thus falling back onto the deterministic case, or on modelling the mathematical and statistical properties of the process, and potentially infering them from the process realisations. The study of the properties of stochastic processes is thus a crucial task in many fields of physics, astronomy, quantitative biology, as well as engineering and finance. Among the classes of stochastic processes, a key role is played by \textit{wide-sense} stationary processes. These are stochastic processes that display an invariance of their statistical properties, such as their two-point autocovariance function, with respect to translation of the independent variable, usually the time $t$. If $x(t)$ is a wide-sense stationary process, its statistical properties are completely determined by the knowledge of the (many-points) autocorrelation functions. In practice, one often has easy access to the two-point correlation function 
\begin{equation}
	C(\tau) = \mathbf{E}[x_t \cdot x_{t+\tau}]
\end{equation}
or, equivalently, to the process \emph{power spectral density} (PSD) $S(f)$. Thanks to the Wiener-Khinchin theorem, in wide-sense stationary processes, the two are in fact related by a Fourier transform: 
\begin{equation}\label{eq:psd-autocorrelation}
	S(f) = \int_{-\infty}^{\infty} \textrm{d}\tau C(\tau) e^{-i 2 \pi f \tau}\,.
\end{equation}
In the context of gravitational waves physics, e.g. \cite{Finn_1992}, the PSD is introduced as  
\begin{equation}\label{eq:psd-f-definition}
	\mathbf{E}[\tilde{x}(f) \cdot \tilde{x}(f^\prime)] = S(f) \delta(f-f^\prime)
\end{equation}
without highlighting its connection with the time structure of the process itself, thus masking some important properties that will be explored further in what follows. The latter definition in Eq. ~\eqref{eq:psd-f-definition} gives, however, i) a straightforward interpretation of the PSD: it measures how much signal ``power" is located in each frequency; ii) an operative way of estimating it for an unknown process.  

An ubiquitous method for such a computation is due to \cite{Welch1967} and it is based on Eqs. (\ref{eq:psd-autocorrelation}-\ref{eq:psd-f-definition}).
The PSD is obtained by slicing the observed realisation $x(t_1),\ldots,x(t_n)$ of the process $x(t)$ into many window-corrected batches and averaging the squared moduli of their Fourier transforms.
This approach is equivalent~\cite{Lomb}\cite{Scargle} to taking the Fourier Transform of the windowed sample autocorrelation $\rho_W$, written as
\begin{equation}
    \rho_{W} = \left\{W_0\rho_0,W_{\pm 1}\rho_{\pm 1}, \dots, W_{\pm M}\rho_{\pm M}, 0, 0, \dots \right\},
\end{equation}
where $\rho$ is the empirical autocorrelation and $M$ is the maximum time lag at which the autocorrelation is computed.
The sequence $W$ is a window function that can be chosen in several different ways, each choice presenting advantages and disadvantages for the final estimate of the PSD.

The choice of a window function is arbitrary and typically is made by trial and error, until a satisfactory compromise between variance and resolution of the estimate of PSD is reached. A high frequency resolution implies high variance and vice-versa.
Besides the window function, Welch's method requires a number of arbitrary choices to be made, such as the number of time slices and the overlap between consecutive slices. All these knobs must be tuned by hand and their choice can dramatically affect the PSD estimation, hence begging the question of what the ``best" PSD estimate is.

Another drawback of this approach is the requirement for the window to be $0$ outside the interval in which the autocorrelation is computed.
We are arbitrarily assuming $\rho_j = 0$ for $j > M$ and modifying the estimate (i.e. the data) if a non-rectangular window is chosen.
Making assumptions on unobserved data and modifying the ones we have at our disposal introduces ``spurious" information about the process that we, in general, do not really have.

A alternative approach providing a smooth PSD estimation, is to adopt a parametric model for the PSD and to fit its parameters to the data with a Reversible Jump Markov Chain Monte Carlo \cite{Cornish_2015,Littenberg_2015}. Despite being effective, this method is problem dependent, since it needs to make definite assumptions on the shape of the PSD. Moreover, it can be computationally expensive and it does not come with a handy implementation available to the public. For all the above reasons, we did not consider such methods in our work.

An appealing alternative, based on the Maximum Entropy principle \cite{JaynesArticle, jaynes2003ptl, Jaynes_MAXENT}, has been derived by~\cite{burg1975maximum}. Being rooted on solid theoretical foundations, we will see that Burg's method, unlike Welch's, does not require any preprocessing of the data and requires very little tuning of the algorithm parameters, since it provides an iterative closed form expression for the spectrum of a stochastic stationary time series. Furthermore, it embeds the PSD estimation problem into an elegant theoretical framework and makes minimal assumptions on the nature of the data.
Lastly and most importantly, it provides a robust link between spectral density estimation and the field of autoregressive processes. This provides a natural and simple machinery to forecast a time series, thus predicting future observations based on previous ones.

In this paper, we discuss the details of the Maximum entropy principle, its application to the problem of PSD estimation with Burg's algorithm and the link between Burg's algorithm and autoregressive process.
Our goal is to bring (again) to public attention Maximum Entropy Spectral analysis, in the hope that it will be widely employed as a way out of the many undesired aspects of the Welch's algorithm (or other similar methods).
To facilitate this goal, we based this study on \texttt{memspectrum}, a freely available, robust and easy-to-use python implementation of the algorithm described below\footnote{
It is available at link: \url{https://pypi.org/project/memspectrum/}.}.
We provide a thorough assessment of the performance of our code and we validate our results performing a number of tests on simulated and real data.
We also compare our results with those of spectral analysis carried out with the standard Welch's method.
In order to apply our model on a realistic setting, we analyse some time series of broad interest in the scientific community.

Our paper is organized as follows: we begin by briefly reviewing the theoretical foundations of the maximum entropy principle in Sec.~\ref{sec:foundations}. Sec.~\ref{sec:validation} presents the validation of Burg's method as well as of our implementation on simulated data. In Sec.~\ref{sec:Welch_comparison} we compare the results from \texttt{memspectrum} with the Welch method; Sec.~\ref{sec:applications} presents a few applications to real time series, including the analysis of GW150914, and, finally, we conclude with a discussion in Sec.~\ref{sec:conclusion}.

\section{Theoretical foundations}\label{sec:foundations}
The Maximum Entropy principle (MAXENT) is among the most important results in probability theory. It provides a way to uniquely assign probabilities to a phenomenon in a way that best represent our state of knowledge, while being non-committal to unavailable information. Its domain of application turned out to be wider than expected. In fact, thanks to~\cite{burg1975maximum}, this method has also been applied to perform high quality computation of power spectral densities of time series.

After a short introduction to Jaynes' MAXENT (Sec.~\ref{sec:MAXENT}), we will review in detail Burg's technique of Maximum Entropy Spectral Analysis (MESA) and show that the estimate can always be expressed in an analytical closed form (Sec.~\ref{sec:MESA}).
Next, we will discuss the interesting link between Burg's method and autoregressive processes (Sec.~\ref{sec:autoregr}) and in Sec.~\ref{sec:forecasting} we will use such link to forecast a time series.

\subsection{Maximum Entropy Principle} \label{sec:MAXENT}

Before introducing the MAXENT principle, we will define, via some simple examples, the two core concepts of the problem and the roles they play in deductive inference: the `evidence' and the `information'.
Let us start with the `information' (or information entropy): it is a measure of the degree of uncertainty on the outcomes of some experiment and specifies the length of the message necessary to provide a full description of the system under study. As an example, no information is brought if we are studying a system whose outcome is certain (the outcome is known with probability $p = 1$), as in this case, a communication is not even needed.
\cite{Shannon} proposed the quantity
\begin{equation}\label{eq:information}
    I = \log_2 \frac{1}{p(x)}
\end{equation}
to measure the quantity of information brought by an outcome $x$ with probability $p(x)$. It is additive quantity as well as a monotonically decreasing function of $p \in [0, 1]$: the more uncertain the outcome, the higher the information it brings.

We can generalize the definition of information in the case where two different outcomes $E_1, E_2$, with given probabilities $P_1$ and $P_2$, are possible.
To gain some intuition on the problem, we ask ourselves which are the probability assignments that make the outcome more uncertain (i.e. maximize the information).
If $P_1$ and $P_2$ are largely different, for instance $P_1 = 0.999$ and $P_2 = 0.001$, we are allowed to believe that event $E_1$ will occur almost certainly, considering $E_2$ to be a very implausible outcome. The information content will be very low.
On the other hand, most unpredictable situation happens when 
\begin{equation}\nonumber
    P_1 = P_2 = \frac{1}{2}:
\end{equation}
this describes a situation of `maximum ignorance' and the information content of such system must be high.
Any generalization of Eq.~\eqref{eq:information}, must then have its maximum when $ P_1 = P_2$.
For $N$ events, the system with the highest possible information content is when:
\begin{equation}\nonumber
    P_1 = \hdots = P_N = \frac{1}{N}:
\end{equation}

\cite{Shannon} showed that the only functional form satisfying continuity with respect to its parameters, additivity and that has a maximum for equal probability events is:
\begin{equation}\label{eq:entropy}
    H[p_1, \dots, p_N] = - \sum_{i = 1}^N p_i\log{p_i}\,,
\end{equation}
which can be interpreted as the `expected information' brought by an experiment with $N$ possible outcomes each with its own probability $p_i$.
In the continuous case:
\begin{equation} \label{eq:entropy_continuos}
    H[p(x)] = - \int p(x)\ln p(x) dx,
\end{equation}
We call the functional $H$ information entropy\footnote{In defining the information entropy as in Eq.~\eqref{eq:entropy_continuos}, we are implicitly assuming a uniform measure over the parameter space. In case of a non-uniform measure $m(x)$, the definition generalises to $ H[p(x)] = - \int p(x)\ln \frac{p(x)}{m(x)} dx$.}.

We now turn to the core of our problem: how can we assign probabilities to a set of events keeping into account our knowledge of the system and, at the same time, ensure it is non-committal towards unavailable knowledge?
The ``knowledge'' at our disposal about the system under investigation is what we define `evidence' and any probability assignment is given such evidence, in agreement with \cite{Cox} construction of probability. In the case above, our knowledge on the system is only the total number $N$ of different outcomes -- this is a minimal requirement. Of course, more complex evidence constraints can be applied.

It is very common that the constraints provided by the evidence are not enough for setting the probabilities for each event: in this case, it is reasonable to assume that the probability assignment should make the experiment as unpredictable as possible\footnote{
In~\cite{Jaynes_MAXENT} this statement is made more precise and justified more thoroughly, with arguments based on combinatorial analysis.
}.
In other words, the information entropy content introduced by the probability assignment should be as large as possible, in accordance with the available evidence. MAXENT formalises this reasoning by stating that probabilities should be assigned by maximizing uncertainty (information entropy) using evidence as a constraint. 
This defines a variational problem, where the information entropy functional $H\left[p_1, \dots, p_N\right]$, defined in Eq. ~\eqref{eq:entropy}, has to be maximized. 

The maximisation of the entropy, supplemented by evidence in the form of constraints to which the sought-for probability distribution must obey, gives rise to several of the most common probability distributions commonly employed in statistics. In the cases of interest, evidence is used to constraint, via Lagrange Multipliers, the momenta of the probability distribuiton we are seeking to evaluate. For instance, whenever the only constraint available is the normalization of the probability distribution (i.e. no evidence is available), the entropy is maximised by the uniform distribution. If we have evidence to constraint the expected value, the information entropy is maximised by the exponential distribution.

Of particular relevance for our purposes is the case in which, in addition to the mean, also the variance is known: MAXENT leads to the Gaussian distribution. 
This derivation is particularly interesting from the foundational point of view, since it provides a deeper insight into the ubiquitous Gaussian distribution.
Indeed, it is not only the limit distribution provided by the central limit theorem for finite variance processes but it is also the distribution that maximizes the entropy for a fixed mean and variance: from the MAXENT principle, it is the correct probability distribution to assign if the mean and covariance are the only quantities that fully define our process. In some sense, we can interpret the central limit theorem as the natural `statistical' evolution toward a configuration that maximizes entropy in repeated experiments.

For this work, we are especially interested in the multi-dimensional case. Suppose we have a vector of measurements $(x(t_1),\ldots,x(t_n)) = (x_1, \ldots, x_n)$ that we conveniently express as a single realization of an unknown stochastic process $x(t)$ and we have information about the expectation value of the process $\mu(t)$ and on the matrix of autocovariances $C_{ij} \equiv C(t_i,t_j)$, then the MAXENT distribution is the $n$-dimensional multivariate Gaussian distribution~\cite{gregory_2005}: 
\begin{align}
    p&\left((x_1, \ldots, x_n)\vert I\right) = \nonumber \\
    &\frac{1}{\left(2 \pi \det C\right)^{k / 2}}\exp\left(-\frac{1}{2}\sum_{i,j}(x_i-\mu_i) (x_j-\mu_j)C^{-1}_{ij} \right)\,. 
\end{align}

For a wide-sense stationary process the mean function is independent of time, hence it can be redefined to be equal to zero without loss of generality, and the auto-covariance function is dependent only on the time lag $\tau \equiv t_i - t_j$. One can thus choose a sampling rate $\Delta t$ so that $C_{ij} = C((i-j)\Delta t)$. The autocovariance matrix thus becomes a Toeplitz matrix\footnote{
We remind the reader that a Toeplitz matrix is a matrix in the form:
$\begin{pmatrix}
	a_0 & a_1 & a_2 & \ldots & \ldots& \ldots &a_n\\
	a_{-1} & a_0 & a_1 & \ldots & \ldots& \ldots &a_{n-1}\\
	a_{-2} & a_{-1} & a_0 & \ldots & \ldots & \ldots  &a_{n-2}\\
	\vdots & \vdots & \vdots & \vdots & \vdots & \vdots &\vdots\\
	a_{-n +1} & \ldots & \ldots & \ldots& a_{-1} & a_0    &a_{1}\\
	a_{-n} & \ldots & \ldots & \ldots& a_{-2} & a_{-1} & a_0
\end{pmatrix}$
}.
Toeplitz matrices are asymptotically equivalent to circulant matrices and thus diagonalized by the discrete Fourier transform base~\cite{Gray}.
Some simple algebra shows that the time-domain multivariate Gaussian can be transformed into the equivalent frequency domain 
probability distribution:
\begin{align}\label{eq:Whittle}
p&\left((\tilde{x}_1, \ldots, \tilde{x}_{n/2})\vert I\right) = \nonumber \\
    &\frac{1}{\left(2 \pi \det S\right)^{n / 2}}\exp\left(-\frac{1}{2}\sum_{ij}\tilde{x}_i S^{-1}_{ij} \tilde{x}_j \right)\,,
\end{align}
where the matrix $S_{ij} = S_i \delta_{ij}$ is an $n\times n$ diagonal matrix whose elements are the PSD $S(f)$ calculated at frequency $f_i$.
Many readers will recognise the familiar form of the Whittle likelihood that stands at the basis of the \emph{matched filter} method \cite{prob_information_theory}
and of gravitational waves data analysis, \cite[e.g.]{Finn_1992,FINDCHIRP}.
Thanks to MAXENT, the problem of defining the probability distribution describing a wide-sense stationary process is thus 
entirely reduced to the estimation of the PSD or, equivalently, the autocovariance function.

\subsection{Maximum Entropy Spectral Analysis} \label{sec:MESA}

In principle, if the autocorrelation was known exactly (i.e. at every time $\tau \in (-\infty,+\infty)$), the computation of the PSD would reduce to a single Fourier transform (i.e. Eq.~\eqref{eq:psd-autocorrelation}). However, in any realistic setting, we are dealing with a finite number of samples $N$ from the process. In such cases, the single periodogram is not a consistent estimator for the power spectral density, since its variance doesn't decrease when the sample size increases. 
Moreover, the error $\sigma_k$ in the estimate of the autocorrelation after $k$ steps increases as $\sigma \sim 1/\sqrt{N - k}$\footnote{
This is easily understood: when computing the autocorrelation at order $k$, only $N-k$ examples of the product $x_t x_{t+k}$ are available and the variance of the average value goes as the inverse of the square root of the points considered.
}, so that only few values for the autocorrelation function can actually be computed reliably.
This bring us to the core of the problem: how to give an estimate from partial (and noisy) knowledge of the autocorrelation function? MAXENT can guide us in this task without any a priori assumptions on the unavailable data\footnote{Indeed this is the largest difference with the most common Welch method. The latter assumes that the unknown values of the autocorrelation are $0$. Clearly, this assumption is unjustified and MAXENT is a good way to relax this assumption.}.

As in the previous examples, one needs to set up a variational problem where the entropy, Eq.~\eqref{eq:entropy_continuos}, is maximized 
subject to some problem-specific constraints. In our case, they are i) the PSD estimate has to be non-negative; ii) its Fourier transform has to match the sample autocorrelation (wherever an estimate of this is available).

Before doing so, there is a technicality to solve: the definition of entropy depends on a probability distribution, not on the PSD.
It can be shown ~\cite[e.g.]{AblesMESA,Bartlett} that the variational problem can be formulated in terms of the power spectral density $S(f)$ alone by considering our signal as the result of the filtering a white noise process using a filter with transfer function $T(f)$ equal to $S(f)$\footnote{
A filter with transfer function $T(f)$ takes in input a time series $x_t$ and outputs a times series $y_t$ such that:
$$T(f) = \frac{\tilde{y}(f)}{\tilde{x}(f)}$$
where $\tilde{x}(f)$ denotes the Fourier transform of $x_t$ (and similarly for $y_t$)
}.
The difference in entropy between the input and the output time series (i.e. the entropy gain) obtained by such filter applied on white noise is:
\begin{equation}\label{eq:EntropyGain}
    \Delta H = \int_{- Ny}^{Ny}\log S(f) df\,.
\end{equation}
where $\Delta t$ is sampling rate and $Ny \equiv \frac{1}{2 \Delta t}$  is the Nyquist frequency.
Thus maximising Eq.~\eqref{eq:EntropyGain} is equivalent to maximizing Eq.~\eqref{eq:entropy_continuos}.

Before maximizing the entropy gain, we need to include the evidence available as a form of mathematical constraints for the assignment of $S(f)$.
This is equivalent in imposing that the variational solution $S(f)$ for the PSD matches the empirical autocorrelation.
Let us define a realization of a stochastic process $(x_1,\ldots,x_N)$ with sample autocorrelations $\bar r_k,\,k=0,\ldots, N/2$, then the PSD must satisfy the following equation:
\begin{equation}\label{eq:MaxConstraint}
\int_{-Ny}^{Ny} S(f) e^{\imath 2 \pi f k \Delta t} df = \bar r_{k}\,.
\end{equation}

Thus, by maximizing Eq.~\eqref{eq:EntropyGain} with constraints in Eq.~\eqref{eq:MaxConstraint}, we can give an estimate of the spectrum given a time series sample.
This approach on PSD computation provides a result consistent with the empirical autocorrelation function whenever this is available and, at the same time, it does not make any assumption for the unavailable estimates for the autocorrelation at large time lags.

Remarkably, the variational problem admits a closed-form analytical expression for $S(f)$.
The expression was first found by ~\cite{burg1975maximum}:
\begin{equation}\label{eq:MESApsd}
    S(f) = \frac{P_N \Delta t}{\left(\sum_{s=0}^{N} a_s z^s\right)\left(\sum_{s = 0}^N a^*_s z^{-s}\right)}\,,
\end{equation}
where $\Delta t$ is the sampling interval of the time series, $z=\exp{(2\pi i f\Delta t)}$, $a_0$ = 1.
The vector obtained as $(1, a_1, \dots, a_N)$ is also known as the \textit{prediction error filter}.
The coefficients $a_s (s > 0)$, together with an overall multiplicative scale factor $P_N$, are to be determined by an iterative process (called Burg's algorithm)
At least, two implementations of Burg's algorithm are available in the literature, labeled as `Standard' and `Fast' in the \texttt{memspectrum} package. The `Standard' method is slower but more stable, while `Fast' trades stability for speed.
On simulated stationary data, both versions typically yield similar results, while our tests with real gravitational waves data seems to indicate that the `Fast' implementation introduces noise into the PSD estimate. \footnote{For this reason, it is advisable to use the `Standard' implementation whenever possible. In most case of numerical instability in the `Fast' method, \texttt{memspectrum} will send a warning to user.}. A comparison of the computational times for Standard MESA implementation and Fast implementation (together with Welch's) is porvided in appendix \ref{sec:computationaltimes}.

The number $N$ of such coefficients is a choice that shall be made by the user and indeed it is the only hyperparameter that needs to be tuned. The details of the derivation and the actual form for the coefficients $a_s$ can be found in Appendix \ref{sec:appendix}.

\subsection{Autoregressive Process Analogy} \label{sec:autoregr}

The application of MESA is not limited to spectral estimates, but it also provides a link between spectral analysis and the study
of autoregressive processes (AR)~\cite{doi:10.1029/RG013i001p00183}.
An autoregressive stationary process of order $p$, AR($p$), is a time series whose values satisfy the following expression: 
\begin{equation} \label{eq:AR_p}
    x_t - b_1 x_{t-1} - b_2 x_{t-2} \dots b_p x_{t - p} = \nu_t
\end{equation}
where $b_1, \ldots, b_p$ are real coefficients and $\nu_t$ is white noise with a given variance $\sigma^2$.
Thus, an AR($p$) process models the dependence of the value of the process at time $t$ from the last $p$ observations, 
thus being potentially able to model complex autocorrelation structures within observations.

Thanks to Wold's theorem~\cite{Wold_theorem}, every stationary time series can be represented as an autoregressive process: this ensures that maximum entropy estimation is faithful and general; it turns out that the maximum entropy principle provides a representation of the time series as an $AR(p)$ process and Burg's algorithm computes the corresponding autoregressive coefficients that are suitable to model the available data.

To show the analogy, we compute the PSD $S_{AR(p)}$ of an $AR(p)$ process and we show that it is formally equivalent to the PSD obtained in Eq.~\eqref{eq:MESApsd}. This will also provide a direct expression for the autoregressive coefficients $b_i$ and for the noise variance $\sigma^2$.
We start taking the $z$ transform~\footnote{The $z$ transform is the discrete-time equivalent of the Laplace transform, thus taking a discrete time-series and returning a complex frequency series.} of Eq.~\eqref{eq:AR_p}: 
\begin{align}
    \sum_t x_t z^t - \sum_i b_i z^i\sum_t x_{t - i} z^{t - i} = \sum_t \nu_t z^t\,.
\end{align}
Calling $\tilde x(z)$ and $\tilde \nu (z)$, the transformed quantities, 
in the $z$ domain, the process takes the form:
\begin{equation}
    \tilde x(z) = \frac{\tilde\nu(z)}{\left(1 - \sum_{n = 1}^p b_n z^n \right)}\,.
\end{equation}
Since we assumed a wide-sense stationary process, $\tilde{x}(z)$ is analytic both on and inside the unit circle. Taking its square value and evaluating it on the unit circle $z = e^{-\imath 2 \pi f \Delta t}$, from the definition of spectral density one obtains:
\begin{equation}\label{eq:ARspectrum}
    S_{AR(p)}(f) = \vert \tilde x(z)\vert ^ 2 = 
    \frac{\vert \tilde \nu(f) \vert ^ 2}{\left\vert 1 - \sum_{n = 1}^p b_n e^{\imath 2 \pi f n \Delta t} \right\vert ^ 2}\,.
\end{equation}
The numerator is the spectral density of white noise $\nu_t$, i.e. its (constant) variance $\sigma^2$.

Eqs.~\eqref{eq:ARspectrum} and ~\eqref{eq:MESApsd} are equivalent, if we identify $b_i = - a_i$ and $P_N \Delta t= \sigma ^ 2$.
This shows that the MAXENT estimation of the PSD models the observed times series as an AR process and provides a {\it fit} for the autoregressive coefficients.
Furthermore, as a consequence of Wold's theorem, there is the theoretical guarantee that every stationary time series can be modelled faithfully by the MAXENT.

\subsection{Forecasting} \label{sec:forecasting}
The link between MESA and AR processes is of particular interest. Given the solution to Burg's recursion to determine the $a_k$, we automatically obtain the coefficients of the equivalent AR process, hence we are able to exploit Eq.~\ref{eq:AR_p} to perform \emph{forecasting}, thus providing plausible future observations, conditioned on the observed data.
Indeed, for an AR($p$) process the conditional probability $p(x_t|x_{t-1}, \ldots , x_{t-p})$ of the observation at time $t$ with respect to the past $p$ observation has the form:
\begin{align}\label{eq:p_forecast}
	p&(x_t|x_{t-1}, \ldots , x_{t-p}) \nonumber\\
	&= \frac{1}{\sigma\sqrt{2\pi}} \exp\left[-\frac{1}{2} \left(\frac{x_t - \sum_{i = 1}^p b_i x_{t-i}}{\sigma}\right)^2\right]\,.
\end{align}
The interpretation of Eq.~(\ref{eq:p_forecast}) is straightforward: $x_t$ follows a Gaussian distribution with a fixed variance and a mean value $m_t = \sum_{i = 1}^p b_i x_{t-i}$ computed from past observations.
Eq.~(\ref{eq:p_forecast}) provides then a well defined probability framework for predicting future observations: this is a very useful feature of MESA, that does not have an equivalent in any other spectral analysis computation methods.

\subsection{Whitening} \label{sec:whitening}

The theory of the AR processes can be also applied to the problem of whitening a time series.
Given a time series, $x_t$, the whitening operation produces another time series $x^W_t$ such that:
\begin{equation}\label{eq:whitening_definition}
	x^W_t = \mathcal{F}^{-1}\left[ \frac{\tilde{x}(f)}{\sqrt{S(f)}} \right]
\end{equation}
where $\mathcal{F}^{-1}$ denotes the inverse Fourier transform of a frequency series.
If $x_t$ is a realization of gaussian noise (see Eq.~\eqref{eq:Whittle}) with PSD $S(F)$, the whitened time series $x^W_t$ is just white noise (i.e. uncorrelated samples from a normal gaussian).

From Eq.~\eqref{eq:AR_p}, remembering that $b_i = - a_i$, it's straightforward to derive an expression for the whitened time series $x^W_t$:
\begin{equation}
	x^W_t = \frac{1}{\sqrt{P_N}} \sum_{i=0}^p a_i x_{t-i}
\end{equation}
This amounts to a convolution of the time series  $x_t$ with the kernel $(1, a_1, \hdots, a_p)$, plus a variance rescaling.
Performing a convolution is an appealing alternative to evaluating Eq.~\eqref{eq:whitening_definition} directly.

\section{Validation of the model}\label{sec:validation}

MESA provides a recursive formula for computing the coefficients $a_k$ in Eq.~(\ref{eq:MESApsd}). The number $M$ of such coefficients is equivalent to the maximum order of the autocorrelation $\bar{r}_m$ considered. In an ideal scenario, this would be equal to the number of points the autocorrelation is computed at (equivalent to the length of data considered). However, the computation of high order coefficients of the autocorrelation is unstable and for high enough $m$, as the estimation for  $\bar{r}_m$ shows a very high variance, broadly scaling as $\sim \left(\sqrt{M - m}\right)^{-1}$.

It is then clear that the choice of the number of samples of the discrete autocorrelation to consider is important: 
on the one hand it is advisable to include as much knowledge of the autocorrelation as possible, leading to include all the known $\bar{r}_m$; on the other hand, including values of the autocorrelation that are not reliably estimated, can be counterproductive.
The order $M$ of the autocorrelation to be considered (or, equivalently, the order $M$ of the underlying autoregressive process) is the only tuning parameter of MESA and a careful balance between these two necessities must be made when applying the algorithm.

The remainder of this section is devoted to an extensive study on how to make such choice.
In Section~\ref{sec:optimizers}, we are going to define two different \textit{loss functions} to measure how well the 
algorithm is able to reproduce a known PSD.
The basic idea is to validate, as the autoregressive order considered increases, the performance of the algorithm results 
by measuring the loss function and pick, among the orders the one that yields better results.
The performance of the different losses will be assessed by answering to two questions: (i) how well the AR order is recovered and (ii) how well the measured PSD is able to whiten the input time series.
This will be discussed Sec.~\ref{sec:arp_validation} and Sec.~\ref{sec:whitening_validation}.

\subsection{Choice of the autoregressive order}\label{sec:optimizers} 

Guided from numerical experiments, an indication on the upper bound to the autoregressive order $M_{max}$ is~\cite{doi:10.1190/1.1440902}:
\begin{equation}\label{eq:MMAx}
M_{max} = 2N / \ln{(2N)}\,,
\end{equation}
where $N$ is the number of observed points in the time-series.
However, this is just a plausible upper limit on the order of the AR process $m$ and the optimal algorithm could employ fewer points.
We then need a more sophisticated method for computing the right value for $m$.
We summarise them below:

\begin{itemize}
\item \textbf{Final prediction Error} 
The first criterion is due to ~\cite{Akaike1970StatisticalPI}. It was proposed that $m$ should be chosen as the 
length that minimizes the error when the filter is used as a predictor, the \emph{final prediction} error (FPE): 
\begin{equation}
    FPE(m) = \mathbb{E}\left[ \left((x_t - \hat x_t) ^ 2\right) \right]
\end{equation}
with $\hat{x}_t = \sum_{i = 1}^M a_i x_{t - i}$.
Asymptotically minimizing FPE is equivalent to minimizing the quantity: 
\begin{equation}
    \mathcal{L}_{\rm FPE}(m) = P_{m} \frac{N + m + 1}{N - m - 1}
\end{equation}
with $P_m$ being the estimated noise variance at order $m$, see Eq.~\eqref{eq:errorFilter1}. In the $N \to \infty$ limit, 
remembering $m_{max} \sim 2N / \log(2N)$, Akaike's loss function is equivalent to the minimization of the variance $P_m$ of the white noise of the underlying $AR(p)$ model. 

\item \textbf{Variance Maximum (VM)}
This second criterion \cite{kay1988modern} is based on a similar assumptions to FPE. It minimises the actual value of the least squares (instead of relying to asymptotical behaviour). t the normalising factor takes into account the k degrees of freedom necessary to estimate the forward prediction error filter $a_k$.\\ The quantity to be minimised is
\begin{equation}
VM(m) = \frac{1}{N - 2m}\sum_{t=m}^N\left(x_t - \sum_{i=1}^m a_i x_{t-i} \right)^{2}
\end{equation}
The package implementation of VM loss function takes advantage of a recursive re-writing of the above formula, as in Eqs (27) and (28) of \cite{Cuoco_2001}.  \\ \\

Several other criteria are available in the literature \cite{doi:10.1029/WR018i004p01097,bhansali1986} and some are implemented in the memspectrum package. We don't report them in this paper since they didn't show any additional merit with respect to the aforementioned loss functions \\
\end{itemize}

Once a loss function is selected, the choice of the best recursion order is straightforward: we solve the Levinson recursion~\cite{doi:10.1002/sapm1946251261} until $M_{max}$, as given in Eq.~\eqref{eq:MMAx}, iterations are reached. Then, the order $m$ is selected to be the one that minimizes the specified loss function.

In a real implementation of the algorithm, computing all the recursion up to $M_{max}$ can result in a significant waste of computational power: the optimal value is often $m_{opt} << M_{max}$ and, in such cases, computing all the values of $m$ until $M_{max}$ is not useful.
In practice, we can apply an \textit{early stop} procedure: every few iterations we look for the best order of $m_{opt}$; if this value does not change for a while, we assume that a good (local) minimum of the loss function is found and the computation is stopped.

The following sections will be devoted to the study of the statistical properties of the loss functions introduced above: we need to understand which choice provides the best quality in the reproduction of some known power spectral densities. In the following paragraph, we will discuss three different comparison (one qualitative and two quantitative) of the two proposed loss functions, o 
\subsection{How accurate are the reconstructed PSDs?}\label{sec:psd_validation}

\begin{figure}
	\centering
	\includegraphics[width = \linewidth]{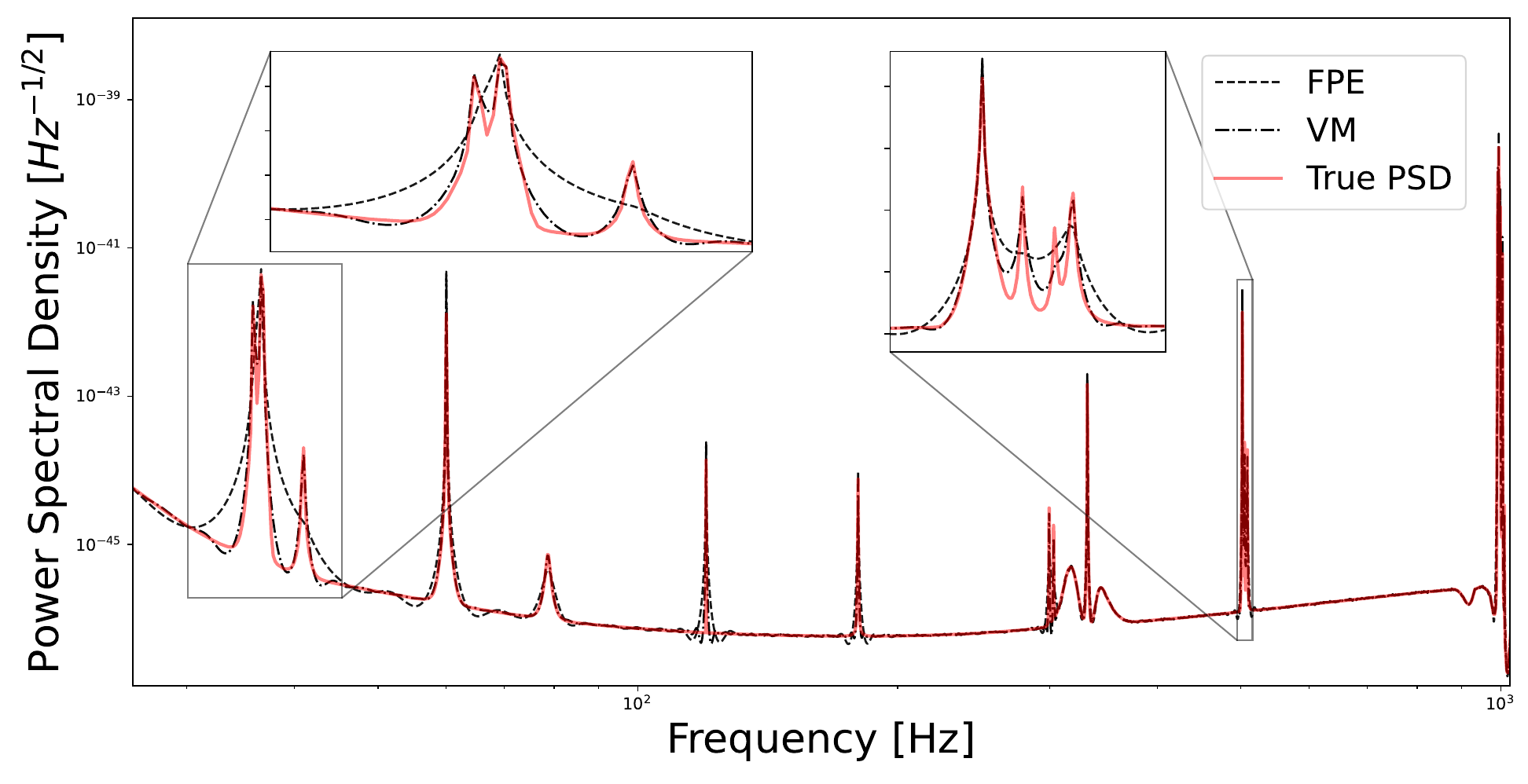}
	\label{fig:psd_comparison}
	\caption{Comparison of the ensemble average PSD estimate}
\end{figure}
In our initial qualitative comparison, depicted in Figure \ref{fig:psd_comparison}, we juxtapose the reconstruction of a known a-priori Power Spectral Density with those obtained using the two distinct loss functions. The red line in the plot represents our chosen reference PSD, derived from the estimated PSD for the GW150914 event.\\ 
To conduct this analysis, we generated 1000 noise time-series whose power spectral densities matches the reference PSD by construction \cite{noiseGen}. The sampling rate and observation time were fixed at $dt = 2048$Hz and $T = 5$s, respectively. For each noise realization, we employed both the FPE and the VM loss functions to estimate the PSD. Ultimately, we compared the reference PSD against the ensemble average of these two estimation methods.\\
The FPE-derived estimate, represented by the dashed line, effectively identifies and reconstructs peaks across both high and low frequency ranges with commendable accuracy. However, as illustrated in the inset plots, FPE struggles when confronted with structured peaks—those containing subordinate modes. In such cases, FPE accurately captures the primary mode but overlooks the subsidiary peaks. \\ 
On the other hand, the VM estimate, depicted as a dot-dashed line, excels in reconstructing both dominant and subordinate modes with remarkable precision. VM appears to prioritize comprehensive mode reconstruction, while FPE emphasize an accurate reconstruction of major modes while potentially neglecting more intricate sub-peaks.

\subsection{How well is the AR order recovered?}\label{sec:arp_validation}

\begin{figure}
	\centering
	\includegraphics[width = \linewidth]{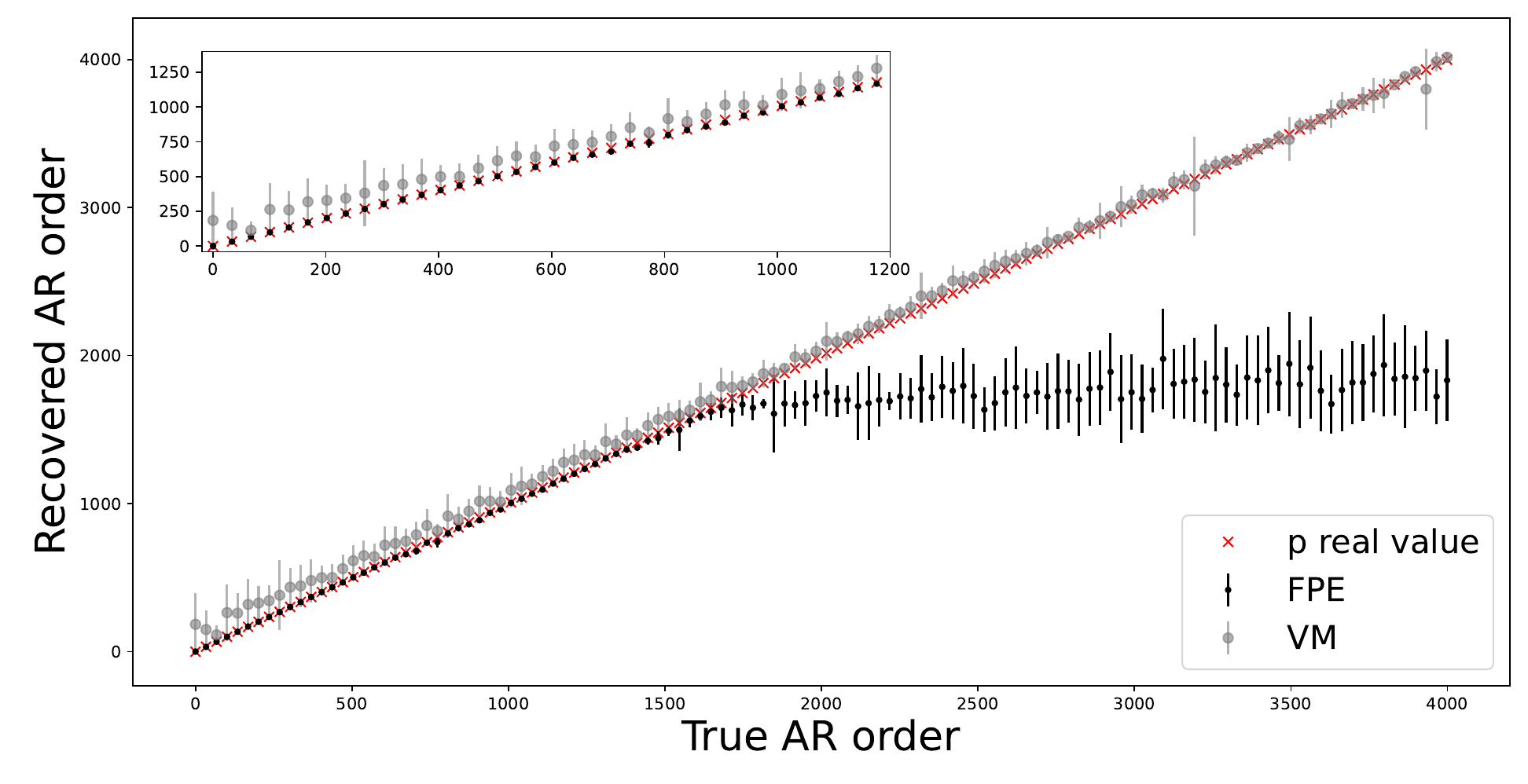}
	\caption{Reconstructed value for the autoregressive order plotted against the true value of the autoregressive order.
	The reconstructed autoregressive orders are computed from a time series randomly drawn with an $AR(p)$ model, with the two different loss functions under investigation.
	}
	\label{fig:p_vs_ptrue}
\end{figure}

Moving to our second comparison, we now focus on another crucial aspect: how accurately each loss function estimates the Autoregressive (AR) order, which represents the number of employed $a_k$ coefficients. \\ 
Here, the memspectrum package proves quite useful. It allows us to assign a specific order to the reconstructed autoregressive filter and use the resulting coefficients to forecast time series. With these tools in hand, we generated various time series, each with a different autoregressive order ranging from m = 0 to m = 4000.\\ 
To ensure reliability, we created 30 distinct time series for each autoregressive order. This approach lets us compute both the mean and variance, giving us insights into the accuracy of each loss function's order estimation. This analysis provides valuable information about how well each method performs in estimating the Autoregressive order across a broad spectrum of scenarios. \\ 
The results are reported in Figure \ref{fig:p_vs_ptrue}. The injected autoregressive order's true value is depicted by the red line. The estimations yielded by the two loss functions are illustrated alongside, accompanied by error bars indicating one standard deviation. \\ 
The plot reveals two distinct regions: one with "short" autoregressive orders (m = 0 to around m = 1600) and another with "long" autoregressive orders (starting from m = 1600).\\ 
In the first region, both loss functions provide comparable results that generally match the actual autoregressive order. FPE performs particularly well, offering estimates close to the injected order and with minimal error bars. VM performs slightly worse than FPE in this range, overestimating complexity and showing larger error bars.\\ 
Moving into the second region ($m > 1600$), a shift in performance becomes apparent. FPE's estimates tend to stabilize at a certain autoregressive value. However, as the injected model becomes more complex beyond this point, FPE's accuracy in recovering the true order diminishes, and its variance increases. In contrast, VM performs better in this range, closely following the actual behavior and consistently recovering the true order within one standard deviation.
To conclude, VM appears to prioritize complexity in its approach. In contrast, FPE seems to lean toward synthesis, emphasizing accurate reconstruction of not too complex models. 

\subsection{How well can MESA whiten the data?}\label{sec:whitening_validation}
\begin{figure}
	\includegraphics[width = \linewidth]{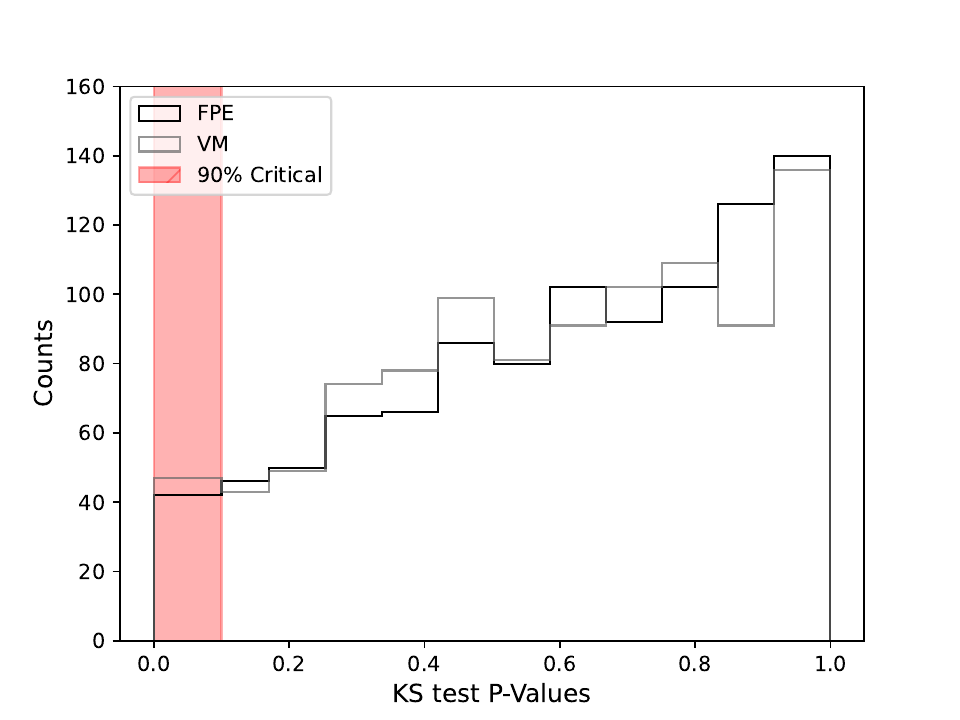}
	\caption{Histogram of the P-values obtained with a Kolmogorv Smirnov test on the whitened time series, against a univariate, 0 mean normal distirbuition}
	\label{fig:whitening}
\end{figure}
In Section \ref{sec:whitening}, we showed how autoregressive coefficients and noise variance estimate $P$ can jointly be used to create a whitening filter, as in Eq. (\ref{eq:whitening_definition}) and . To complete our investigation, we compare how well these whitening filters work when obtained from the two different loss functions we've been studying. \\ 
For this test, we employed the same set of time-series data as described in Section \ref{sec:psd_validation}. Each time series underwent the whitening process using the autoregressive filter derived from its corresponding loss function. We then evaluated the resulting whitened time series against a zero-mean, univariate normal distribution using the Kolmogorov-Smirnov test. \\
The results are reported as an histogram for the obtained p-values in Figure \ref{fig:whitening} together with the chosen critical region of $p < 0.1$, representing a $90\%$  confidence level. In this region, there is no statistical difference between the two. Infact, the total number of counts in this bin are respectively $c_{VM} = 47 \pm 7$ and $c_{FPE} = 42 \pm 6$, affirming the absence of a pronounced discrepancy between the two.
In essence, this final examination underscores a shared proficiency in whitening between the two loss functions, showing that very long filters are not needed to obtain a comparlable result in whitening. Both methods showcase comparable results for whitening scopes.\\

From our previous discussions, it's evident that both FPE and VM have their own strengths, and the choice between them greatly depends on the specific analysis requirements. In our analysis, VM tends to provide more accurate PSD estimates and often results in longer autoregressive filters. However, in cases where the underlying model is simple, there is a risk of VM overestimating complexity and generating patterns that don't truly reflect the data.\\ 
On the other hand, FPE is a good option for reconstructing processes without introducing unnecessary complexity. However, it might underestimate the complexity of the data, particularly in scenarios involving secondary peaks or in the low-frequency region.\\ 

Lastly, it's worth noting that FPE holds the advantage of lower numerical complexity due to its straightforward calculations involving simple arithmetic. In contrast, VM requires more complex computations, dealing with arrays that might be very long depending on the analysed data.

\section{Comparison with Welch method}\label{sec:Welch_comparison}

We perform a {\it qualitative} comparison between the performance of the MESA and of the standard Welch algorithm.
In this, we cannot avoid to be only qualitative. Indeed, as the results of the comparison are problem dependent, it is very hard to quantify this in a single metric. Although similar studies can be drawn from any other PSD, in this section we focus on a single PSD and we try to generalize some observations that we make.
We decide to use the analytical PSD computed for the LIGO Handford interferometer, released together with the GWCT-1 catalog~\cite{GWTC1,PSD_release}, and computed with the BayesLine package \cite{Cornish_2015,Littenberg_2015,Cornish_2020,Chatziioannou_2019}.

\begin{figure*}

\begin{minipage}{0.99\columnwidth}
	\caption{Comparison between analytic (dashed line) and estimated (red line) spectrum. The estimation is performed with Maximum Entropy method on \textit{synthetic} data, with an increasing observation time $T = 1, 5, 10, 100, \SI{1000}{s}$.}
	\label{fig:MESA_LIGO_data}
	\includegraphics{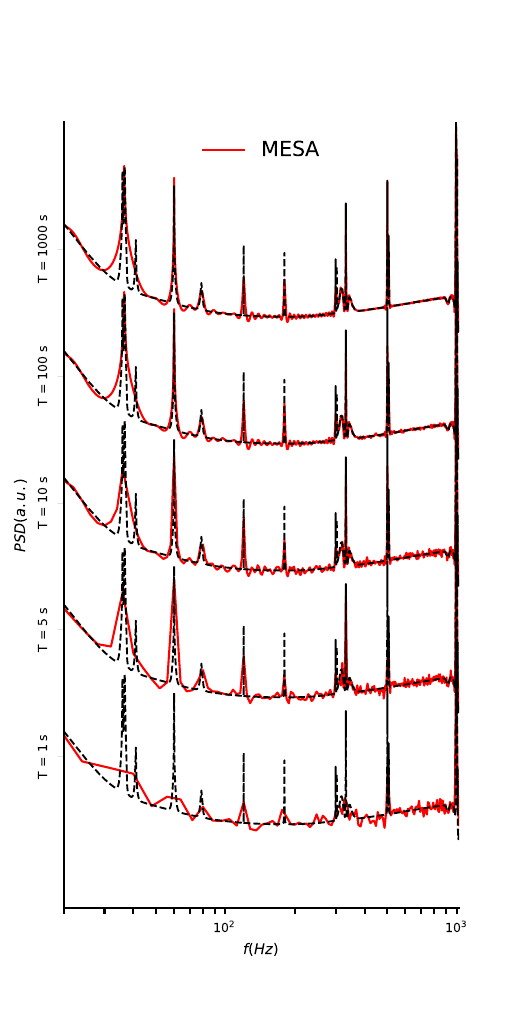}
\end{minipage}\hfill
\begin{minipage}{0.99\columnwidth}
	\caption{Comparison between analytic (dashed line) and estimated (green line) spectrum. The estimation is performed with Welch's method on \textit{synthetic} data with an increasing observation time $T = 1, 5, 10, 100, \SI{1000}{s}$.}
	\label{fig:welch_LIGO_data}
	\includegraphics{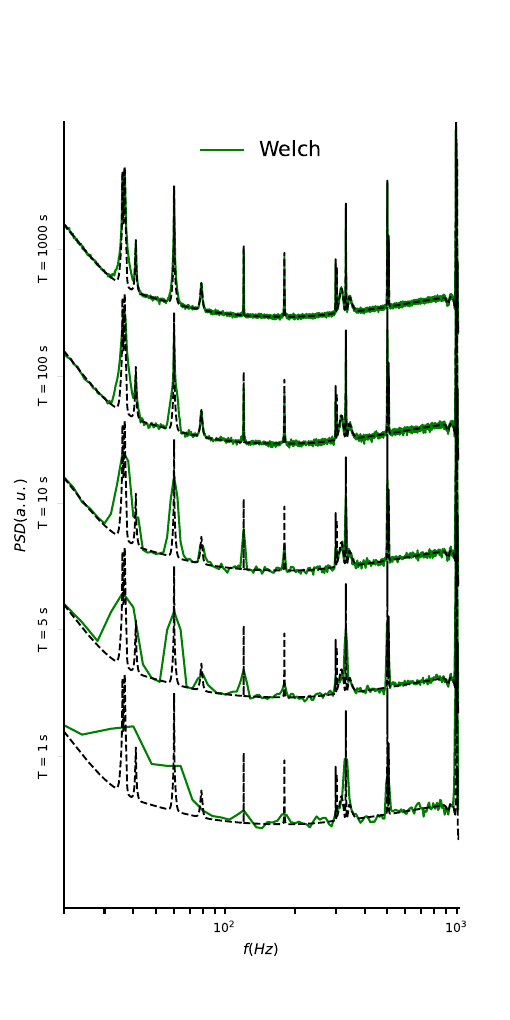}
\end{minipage}

\end{figure*}

We simulate data\footnote{This is to ensure that we have a baseline PSD to compare the data with} from the PSD used for the analysis of the event GW150914 and we employ both Welch's method and MESA to estimate the spectrum.
We vary the length of the data used for the estimation: this is also useful to assess how the computation depends on the data available. We set the total observation time $T = 1, 5, 10, 100, 1000 \SI{}{s}$
For the MESA algorithm, we choose the VM loss function. For the Welch algorithm, we employ a Tukey window with the shape parameter $\alpha$ 
equal to 0.4 (see scipy documentation), an overlap fraction of $1/2$ for the segments and a length of 
segments $L = 512, 1024, 2048, 8192, 32768$ points, depending on the observation time.
In all cases, the sampling rate is set to $\SI{4096}{Hz}$.
For the Welch algorithm, we use the standard implementation provided by the python library \texttt{scipy}~\cite{numpy,scipy}.
The results from both methods are summarized in  Figures ~\ref{fig:MESA_LIGO_data} and~\ref{fig:welch_LIGO_data} respectively.

First of all, we note that using a longer time series results in a better estimation of the PSD, especially at low frequencies. 
This is somehow obvious: longer data streams probe lower frequencies thanks to Nyquist's theorem as well as providing better estimates for the FFT, in the Welch case, and the sample autocorrelation, for MESA.

We also note that MESA converges (Figures ~\ref{fig:MESA_LIGO_data} and~\ref{fig:welch_LIGO_data}) to the underlying spectrum much faster than Welch's method, providing a better estimate even in the case of short time series. Although observed at every frequency, this behaviour is more evident in the low frequency region. An accurate profile reconstruction can be obtained with MESA using a 5 seconds-strain only, while Welch method requires at least 10 seconds of data to obtain a comparable profile.
Furthermore, MESA is able to model all the details of the peak at around $\sim \SI{40}{Hz}$ (even with $T=\SI{100}{s}$), while the Welch's algorithm fails to do so even with an observation time of $T=\SI{1000}{s}$.

Another important element is the noise of the spectral estimation: we find that the PSD estimation provided by the Welch's method is noisier (i.e. has a large number of spurious peaks) compared to the PSD measured with MESA and FPE loss function. This is especially true at high frequencies and for long observation times $T$.

Finally, as already discussed Welch's method is very dependent on the choice of window function. 
A Tukey window with aforementioned parameters is what we found to be the best compromise between noise 
and accuracy for the reconstruction, but different choices can be made, possibly providing more accurate results than the ones reported here. 
However, we want to stress that this fact does not invalidate our discussion but reinforces it: one of the most appealing advantages of MESA is the minimal amount of fine tuning required.

\section{Marginalisation over the noise distribution: application to GW parameter estimation} \label{sec:applications}


Define the data hypothesis $D$ as the statement that the data $D = S + N$ with $S$ and $N$ some deterministic signal and some noise hypotheses. Typically, in this formulation one is choosing both a functional form for the signal of interest $S\equiv ``h(t;\theta)''$ and some parametric form $f(t)$ for the noise distribution $N\equiv ``n(t) \sim f(t)''$. Some well established math then leads to the usual Bayesian framework for parameter estimation, see \cite{lalinference} for an application to gravitational wave physics. This procedure is very robust as long as the choice of noise distribution is indeed representative of the underlying process. Let us relax the $N$ hypothesis by defining a \emph{residuals} hypothesis $R$ as $R = D - S$. This might seem a very trivial statement, but it has a non trivial application: given  $d(t) = h(t;\theta) + n(t)$ where $h(t;\theta)$ is our signal model, defined by a set of parameters $\theta$,  the residuals $r(t) \equiv d(t) - h(t;\theta)$. Formally, no reference to the noise process is present anymore. Under MAXENT, we can model $r(t) \sim \mathrm{AR}(k)$ with $k$ the \emph{unknown} order of the process to be inferred from the residuals, either via one of the aforementioned loss functions or even by marginalising over it while exploring the signal space. Moreover, we can \emph{always} write $p(r(t)|N\,I)$ as in Eq.~(\ref{eq:Whittle}) once we know $k$, with the PSD given in Eq.~(\ref{eq:ARspectrum}), whatever the noise process actually is. In other words, we care only about maximising the information entropy in the distribution of the residuals.

Hence, as an application of MESA, and its implementation in \texttt{memspectrum}, we analyse GW150914\cite{gw150914} using a Bayesian framework that allows for the marginalisation of the order  $k$ of the AR($k$) process representing the residuals data stream. Although the inference is essentially unchanged compared to the standard case, see \cite{lalinference}, there are some substantial modification to the likelihood construction. Since MESA is applicable to time-domain data, all calculations prior to the Fourier transform must be performed in time domain, thus increasing the computational cost by a non-negligible amount. We shall refer the time-of-arrival parameter $t_c$ of the GW to the geocenter. At each iteration of the inference algorithm, we sample a vector $\theta\equiv\theta_{GW}\cup k$\footnote{We indicate the set of all GW parameters (component masses, spins, luminosity distance, etc.) with $\theta_{GW}$.}. For each inteferometer $j$, therefore, we need to compute a time-delay $\Delta t_j$ to compute the antenna response functions $F_{j,+}(t+\Delta t_j),F_{j,\times}(t+\Delta t_j)$ as well as the correct time-shift for the GW template 
\begin{align}
h_j(t) =\sum_{p = +,\times}F_{j,p}(t+\Delta t_j)h_{p}(t+\Delta t_j;\theta_{GW})
\end{align}
that we use to compute the time-domain residuals $r_j(t) = d_j(t) - h_j(t)$. We apply \texttt{memspectrum} to $r_j(t)$ with the \emph{fixed} value of $k$ and calculate the detector likelihood for $\tilde{r}_j$ using Eq.~\eqref{eq:Whittle} and PSD as in Eq.~\eqref{eq:MESApsd}. The coherent likelihood is then given by the product of the individual likelihoods. As our analysis template, we adopt the fast machine learning based MLGW model \cite{MLGW}, an aligned spin model trained on TEOBResumS \cite{teobresums}, that has been shown to perform well on LVK events detected during O1 and O2 \cite{MLGW}. Our sampler is a nested sampling algorithm \cite{john_veitch_2020_4109277} and the specific inference model is implemented as part of \texttt{granite}, a dedicated inference model for ground-based interferometric detectors. We compare our results with the combined posterior samples available from GWOSC \cite{GWOSC} and available at \linebreak
\texttt{https://zenodo.org/records/6513631}.

\begin{figure}
	\caption{Posterior samples for $\mathcal{M}$ and mass ratio $q$ from the LVK (blue) and using \texttt{memspectrum} (red). The samples are largely consistent among the two models, with the MESA model providing a more conservative estimate.}
	\label{fig:gw150914_masses}
	\includegraphics[width=0.5\textwidth,keepaspectratio]{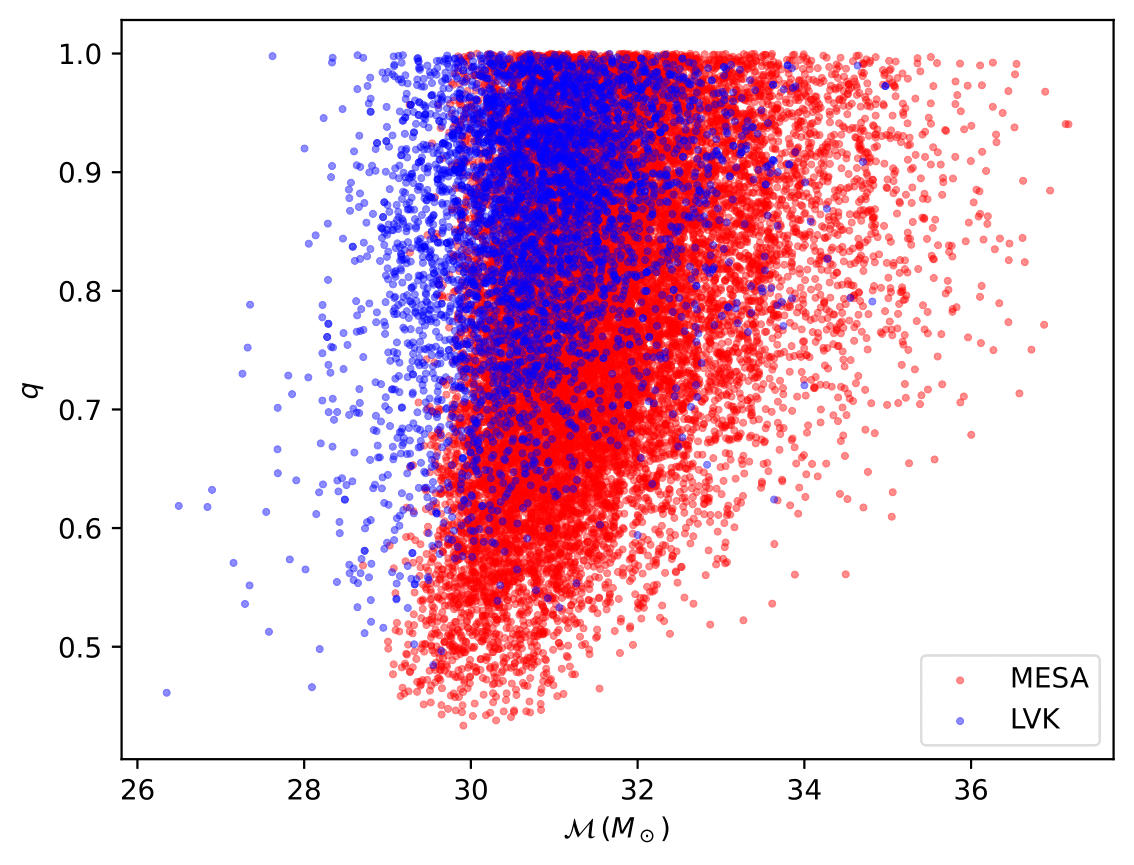}
\end{figure}
\begin{figure}
	\caption{Posterior samples for the sky position angles from the LVK(blue) and using \texttt{memspectrum}(red). The samples are largely consistent among the two models, with the MESA model providing a more conservative estimate.}
	\label{fig:gw150914_sky}
	\includegraphics[width=0.5\textwidth,keepaspectratio]{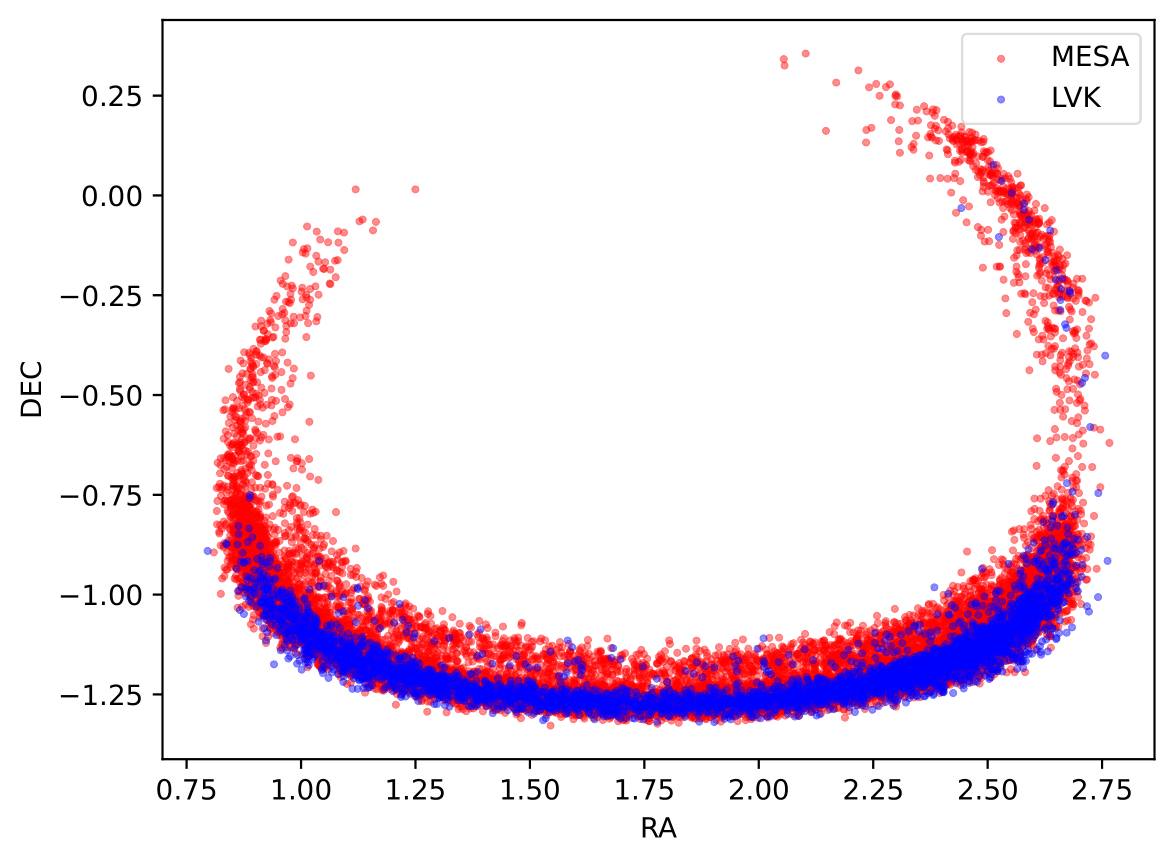}
\end{figure}
\begin{figure}
	\caption{Whitened reconstructed waveforms and data from our analysis for the Hanford detector (top panel) and the Livingston detector (bottom panel). The shaded turquoise area indicates the 90\% credible region over the waveforms space while the purple contours indicate the 90\% credible regions over the whitened data.}
	\label{fig:gw150914_waveforms}
	\includegraphics[width=0.5\textwidth,keepaspectratio]{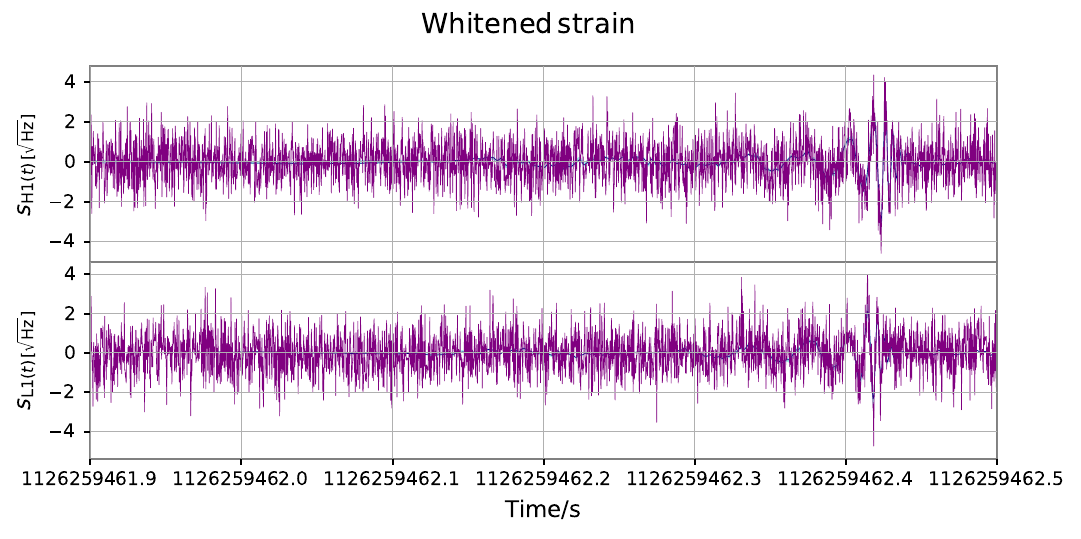}
\end{figure}
\begin{figure}
	\caption{Posterior distributions for AR process orders in the Hanford (red) and Livingston (blue).}
	\label{fig:gw150914_orders}
	\includegraphics[width=0.5\textwidth,keepaspectratio]{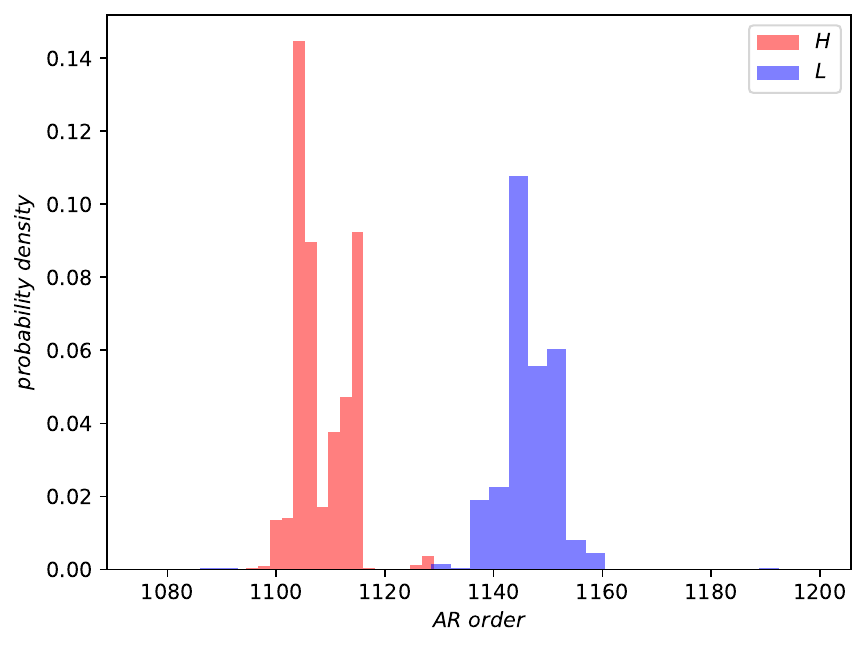}
\end{figure}

\begin{figure}
	\caption{\emph{Top panel}: Posterior PSDs for the Hanford (orange line) and Livingstone (blue line) as inferred by our analysis. \emph{Bottom panel}: Relative uncertainty around the median for both the Hanford(blue) and the Livingston(orange) PSDs.}
	\label{fig:gw150914_psds}
	\includegraphics[width=0.5\textwidth,keepaspectratio]{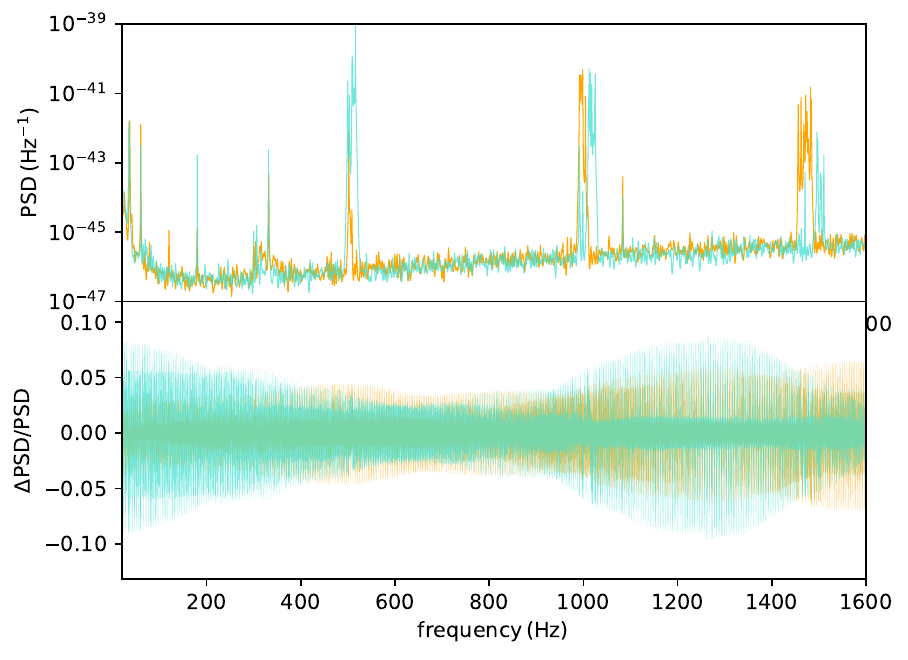}
\end{figure}

In Figs.~\ref{fig:gw150914_masses}, ~\ref{fig:gw150914_sky} and \ref{fig:gw150914_waveforms} we show the posteriors for the set of intrinsic parameters, extrinsic parameters and reconstructed waveform from our analysis. Our results can be summarised as follows: our posterior samples are in general consistent with what has been released by the LVK, however our credible regions tend to be larger. This is expected since our likelihood includes additional uncertainty due to the explicit sampling over the process order, hence the PSD. For the particular 4 seconds of data, sampled at 4096\,Hz, the recovered orders are $k_{H1} = 1107_{-5}^{+9}$ and $k_{L1} = 1146_{-8}^{+8}$, Fig.~\ref{fig:gw150914_orders}. The corresponding PSDs and uncertainties are shown in Fig.~\ref{fig:gw150914_psds}. The full joint posterior distribution recovered when marginalising over the AR order is shown in Appendix~\ref{sec:appendix_b}.

\section{Summary and discussion} \label{sec:conclusion}

We presented a case study of the application of Maximum Entropy principle to the realm of spectral estimation. Albeit the methodology hereby presented is grounded on solid theoretical foundations and its merits are widely recognised, Maximum Entropy methods have yet to be adopted routinely in the study of problems related to time series. The superior nature of maximum entropy methods, and in particular of Burg's method, is exemplified by the closed form estimate of the power spectral density and by the theoretical bridge between spectral analysis and AR processes. Moreover, the method presents, in our view, two main advantages when compared with more traditional ones; first there is no need to choose an arbitrary window function to correct the data and, second it provides as straightforward way to compute predictions given past observations. Accompanying this work, we provide a publicly available \texttt{Python} implementation, called \texttt{memspectrum}, that we used to perform the numerical studies presented in this work.. 

Since the order of the AR process is not yet determined by the theory, we opted for an in-depth investigation of several proposals in the literature and found that different loss functions are required for different situations, with the FPE loss function being the most indicated to deal with gravitational wave data. Along these lines, we directly compared the PSDs computed with MESA with the canonical Welch's algorithm. As outlined in Sec.~\ref{sec:Welch_comparison}, MESA provides PSD estimates with smaller variance and better accuracy than Welch algorithm.
The use of MESA is particularly useful for short time series samples, where Welch's method is outperformed in both precision and confidence. As an examples, Figures ~\ref{fig:MESA_LIGO_data} and~\ref{fig:welch_LIGO_data} illustrate that MESA's performance over a 10-second interval is more closely aligned with Welch's performance over a 100-second interval than Welch's performance over a 10-second interval alone. 

This observation suggests a promising avenue to pursue in future developments of gravitational waves data analysis: for short time series, comparable with the length of binary black hole systems as observed by LIGO, Virgo and KAGRA, the computational cost of MESA is moderate and the inferred PSD is an accurate representation of the true underlying PSD. By applying MESA to 4 seconds of data in correspondence to GW150914, we demonstrated that it is possible to simultaneously estimate the signal and noise parameters, hence effectively marginalise over the noise PSD, without the need to 
\begin{itemize}
\item assume a specific functional form for the PSD;
\item estimate the PSD in an off-source segment of data.
\end{itemize}
Both items are of particular interest for several reasons that we shall discuss in what follows. Several proposals exist in the literature attempt to marginalise over the PSD, mostly using a parametric model for the PSD \cite{Littenberg_2013,Edwards_2015, lalinference}. MAXENT fixes the functional form for us exploiting the correspondence with AR processes, providing a one-parameter family of models that are particularly easy to sample, thus grounding the noise properties marginalisation in solid theoretical foundations and in an easy-to-use numerical implementation. The latter point is also particularly relevant, especially in the context of future GW detectors. Future detectors are in fact expected to be operating in the signal dominated regime, with several sources -- potentially from different classes -- constantly present within the detectors' data streams. In these cases the common procedure of estimating the PSD from off-sources segments is bound to fail and or provide biases inferences. MAXENT and MESA model and are relevant \emph{only for the segment of data under consideration}, and make no assumptions over what is not part of the analysis. We believe, and we will show in a future study, that using MESA can be a natural solution for computing single-source posteriors whenever multiple sources are overlapping. This is possible since, in our formulation, everything that we did not label as \emph{signal} will be part of the residuals, over which we apply MESA. 

Furthermore, MESA provides a simple, but robust and quite accurate, albeit for short times, predictor for the time series. This fact is remarkable and can be used in time series analysis for several purposes. As an example, an anomaly detection pipeline could be built using the forecasts of MESA: the predictions can form a baseline to compare the actual observations with. Whenever the observed data are outside the expectations, an anomaly detection can be claimed. Of course such predictions can be done with a more accurate (perhaps nonlinear) model; however MESA has the advantage of being simple and fast to construct, while providing decent predictions. At the same time, several instruments present gaps in their data stream, for instance LISA is expected to show such gaps~(e.g. \cite{lisa_gaps} and references therein), MESA forecasting capabilities could be used to fill those gaps with predicted data from past observations. In conclusion, we reiterate that MESA is a theoretically sound, computationally feasible and reliable way of studying the properties of stochastic processes and we hope that the investigations presented in this work will further stimulate developments and applications of this method.

\paragraph{Acknowledgments}
We are grateful to S.~Biscoveanu, D.~Laghi, M.~Maugeri, C.~Rossi and S.~Shore for useful comments and discussions.\\
This research has made use of data, software and/or web tools obtained from the Gravitational Wave Open Science Center (\url{https://www.gw-openscience.org/}), a service of LIGO Laboratory, the LIGO Scientific Collaboration and the Virgo Collaboration. LIGO Laboratory and Advanced LIGO are funded by the United States National Science Foundation (NSF) as well as the Science and Technology Facilities Council (STFC) of the United Kingdom, the Max-Planck-Society (MPS), and the State of Niedersachsen/Germany for support of the construction of Advanced LIGO and construction and operation of the GEO600 detector. Additional support for Advanced LIGO was provided by the Australian Research Council. Virgo is funded, through the European Gravitational Observatory (EGO), by the French Centre National de Recherche Scientifique (CNRS), the Italian Istituto Nazionale di Fisica Nucleare (INFN) and the Dutch Nikhef, with contributions by institutions from Belgium, Germany, Greece, Hungary, Ireland, Japan, Monaco, Poland, Portugal, Spain.

\bibliographystyle{spphys2}
\bibliography{Bibliography} 

\begin{thebibliography}{10}
\providecommand{\url}[1]{{#1}}
\providecommand{\urlprefix}{URL }
\expandafter\ifx\csname urlstyle\endcsname\relax
  \providecommand{\doi}[1]{DOI \discretionary{}{}{}#1}\else
  \providecommand{\doi}{DOI \discretionary{}{}{}\begingroup
  \urlstyle{rm}\Url}\fi

\bibitem{Finn_1992}
L.S. Finn, Physical Review D \textbf{46}(12), 5236–5249 (1992).
\newblock \doi{10.1103/physrevd.46.5236}.
\newblock \urlprefix\url{http://dx.doi.org/10.1103/PhysRevD.46.5236}

\bibitem{Welch1967}
P.~Welch, IEEE Transactions on audio and electroacoustics \textbf{15}(2), 70
  (1967)

\bibitem{Lomb}
N.R. {Lomb}, Astrophysics and Space Science \textbf{39}(2), 447 (1976).
\newblock \doi{10.1007/BF00648343}

\bibitem{Scargle}
J.D. {Scargle}, The Astrophysical Journal \textbf{263}, 835 (1982).
\newblock \doi{10.1086/160554}

\bibitem{Cornish_2015}
N.J. Cornish, T.B. Littenberg, Classical and Quantum Gravity \textbf{32}(13),
  135012 (2015).
\newblock \doi{10.1088/0264-9381/32/13/135012}.
\newblock \urlprefix\url{http://dx.doi.org/10.1088/0264-9381/32/13/135012}

\bibitem{Littenberg_2015}
T.B. Littenberg, N.J. Cornish, Physical Review D \textbf{91}(8) (2015).
\newblock \doi{10.1103/physrevd.91.084034}.
\newblock \urlprefix\url{http://dx.doi.org/10.1103/PhysRevD.91.084034}

\bibitem{JaynesArticle}
E.T. Jaynes, Physical Review \textbf{106}, 620 (1957)

\bibitem{jaynes2003ptl}
E.~Jaynes, G.~Bretthorst, \emph{{Probability Theory: The Logic of Science}}
  (Cambridge University Press:, 2003)

\bibitem{Jaynes_MAXENT}
E.T. {Jaynes}, Proceedings of the IEEE \textbf{70}(9), 939 (1982).
\newblock \doi{10.1109/PROC.1982.12425}

\bibitem{burg1975maximum}
J.~Burg, \emph{Maximum Entropy Spectral Analysis}.
\newblock Stanford Exploration Project (Stanford University, 1975).
\newblock \urlprefix\url{https://books.google.it/books?id=Xug\_AAAAIAAJ}

\bibitem{Shannon}
C.E. Shannon, Bell System Technical Journal \textbf{27}(3), 379 (1948)

\bibitem{Cox}
R.T. {Cox}, American Journal of Physics \textbf{14}(1), 1 (1946).
\newblock \doi{10.1119/1.1990764}

\bibitem{gregory_2005}
P.~Gregory, \emph{Multivariate Gaussian from maximum entropy} (Cambridge
  University Press, 2005), p. 450–454.
\newblock \doi{10.1017/CBO9780511791277.020}

\bibitem{Gray}
R.M. Gray, \emph{Toeplitz and Circulant Matrices: A Review} (Now Foundations
  and Trends, 2006).
\newblock \doi{10.1561/0100000006}

\bibitem{prob_information_theory}
D.W.F. P.~M.~Woodward, W.~Higinbotham, \emph{Probability and information
  theory, with applications to radar}, 2nd edn. (Pergamon Press, 1964).
\newblock \urlprefix\url{http://cds.cern.ch/record/2031792}

\bibitem{FINDCHIRP}
B.~Allen, W.G. Anderson, P.R. Brady, et~al., Phys. Rev. D \textbf{85}, 122006
  (2012).
\newblock \doi{10.1103/PhysRevD.85.122006}.
\newblock \urlprefix\url{https://link.aps.org/doi/10.1103/PhysRevD.85.122006}

\bibitem{AblesMESA}
J.G. {Ables}, Astronomy and Astrophysics Supplement \textbf{15}, 383 (1974)

\bibitem{Bartlett}
M.~Bartlett, Louvain Economic Review \textbf{34}(2), 227–227 (1968).
\newblock \doi{10.1017/S077045180004077X}

\bibitem{doi:10.1029/RG013i001p00183}
T.J. Ulrych, T.N. Bishop, Reviews of Geophysics \textbf{13}(1), 183 (1975).
\newblock \doi{10.1029/RG013i001p00183}.
\newblock
  \urlprefix\url{https://agupubs.onlinelibrary.wiley.com/doi/abs/10.1029/RG013i001p00183}

\bibitem{Wold_theorem}
H.~Wold, Journal of the Institute of Actuaries \textbf{70}(1), 113–115
  (1939).
\newblock \doi{10.1017/S0020268100011574}

\bibitem{doi:10.1190/1.1440902}
J.G. Berryman, GEOPHYSICS \textbf{43}(7), 1384 (1978)

\bibitem{Akaike1970StatisticalPI}
H.~Akaike, Annals of the Institute of Statistical Mathematics pp. 137--151
  (1998).
\newblock \doi{10.1007/978-1-4612-1694-0_11}.
\newblock \urlprefix\url{https://doi.org/10.1007/978-1-4612-1694-0_11}

\bibitem{kay1988modern}
S.~Kay, \emph{Modern Spectral Estimation}.
\newblock Prentice-Hall signal processing series (Prentice-Hall, 1988)

\bibitem{Cuoco_2001}
E.~Cuoco, G.~Calamai, L.~Fabbroni, et~al., Classical and Quantum Gravity
  \textbf{18}(9), 1727–1751 (2001).
\newblock \doi{10.1088/0264-9381/18/9/309}.
\newblock \urlprefix\url{http://dx.doi.org/10.1088/0264-9381/18/9/309}

\bibitem{doi:10.1029/WR018i004p01097}
A.R. Rao, R.L. Kashyap, L.~Mao, Water Resources Research \textbf{18}(4), 1097
  (1982).
\newblock \doi{10.1029/WR018i004p01097}.
\newblock
  \urlprefix\url{https://agupubs.onlinelibrary.wiley.com/doi/abs/10.1029/WR018i004p01097}

\bibitem{bhansali1986}
R.J. Bhansali, Annals of Statistics. \textbf{14}(1), 315 (1986).
\newblock \doi{10.1214/aos/1176349858}.
\newblock \urlprefix\url{https://doi.org/10.1214/aos/1176349858}

\bibitem{doi:10.1002/sapm1946251261}
N.~Levinson, Journal of Mathematics and Physics \textbf{25}(1-4), 261 (1946)

\bibitem{noiseGen}
A.J. Owens, Journal of Geophysical Research: Space Physics \textbf{83}(A4),
  1673 (1978)

\bibitem{GWTC1}
B.~Abbott, R.~Abbott, T.~Abbott, et~al., Physical Review X \textbf{9}(3)
  (2019).
\newblock \doi{10.1103/physrevx.9.031040}.
\newblock \urlprefix\url{http://dx.doi.org/10.1103/PhysRevX.9.031040}

\bibitem{PSD_release}
B.~Abbott, R.~Abbott, T.~Abbott, et~al.
\newblock {LIGO Document P1900011-Power Spectral Densities (PSD) release for
  GWTC-1}.
\newblock LIGO Document Service:
  \url{https://dcc.ligo.org/LIGO-P1900011/public} (2019)

\bibitem{Cornish_2020}
N.J. Cornish, T.B. Littenberg, B.~B\'ecsy, et~al., Phys. Rev. D
  \textbf{103}(4), 044006 (2021).
\newblock \doi{10.1103/PhysRevD.103.044006}

\bibitem{Chatziioannou_2019}
K.~Chatziioannou, C.J. Haster, T.B. Littenberg, et~al., Physical Review D
  \textbf{100}(10) (2019).
\newblock \doi{10.1103/physrevd.100.104004}.
\newblock \urlprefix\url{http://dx.doi.org/10.1103/PhysRevD.100.104004}

\bibitem{numpy}
C.R. Harris, K.J. Millman, S.J. van~der Walt, et~al., Nature \textbf{585},
  357–362 (2020).
\newblock \doi{10.1038/s41586-020-2649-2}

\bibitem{scipy}
P.~Virtanen, R.~Gommers, T.E. Oliphant, et~al., Nature Methods \textbf{17}, 261
  (2020).
\newblock \doi{10.1038/s41592-019-0686-2}

\bibitem{lalinference}
J.~Veitch, V.~Raymond, B.~Farr, et~al., Phys. Rev. D \textbf{91}, 042003
  (2015).
\newblock \doi{10.1103/PhysRevD.91.042003}.
\newblock \urlprefix\url{https://link.aps.org/doi/10.1103/PhysRevD.91.042003}

\bibitem{gw150914}
B.~Abbott, R.~Abbott, T.~Abbott, et~al., Physical Review Letters \textbf{116}
  (2016).
\newblock \doi{10.1103/PhysRevLett.116.061102}

\bibitem{MLGW}
S.~{Schmidt}, M.~{Breschi}, R.~{Gamba}, et~al., Phys. Rev. D \textbf{103}(4),
  043020 (2021).
\newblock \doi{10.1103/PhysRevD.103.043020}

\bibitem{teobresums}
A.~{Nagar}, S.~{Bernuzzi}, W.~{Del Pozzo}, et~al., Phys. Rev. D
  \textbf{98}(10), 104052 (2018).
\newblock \doi{10.1103/PhysRevD.98.104052}

\bibitem{john_veitch_2020_4109277}
J.~Veitch, W.D. Pozzo, M.~Williams, et~al.
\newblock johnveitch/cpnest: Fix for python < 3.8 versioning (2020).
\newblock \doi{10.5281/zenodo.4109277}.
\newblock \urlprefix\url{https://doi.org/10.5281/zenodo.4109277}

\bibitem{GWOSC}
R.~{Abbott}, , O.~{Bulashenko}, et~al., Astrophys. J. Suppl. \textbf{267}(2),
  29 (2023).
\newblock \doi{10.3847/1538-4365/acdc9f}

\bibitem{Littenberg_2013}
T.B. Littenberg, M.~Coughlin, B.~Farr, W.M. Farr, Physical Review D
  \textbf{88}(8) (2013).
\newblock \doi{10.1103/physrevd.88.084044}.
\newblock \urlprefix\url{http://dx.doi.org/10.1103/PhysRevD.88.084044}

\bibitem{Edwards_2015}
M.C. Edwards, R.~Meyer, N.~Christensen, Physical Review D \textbf{92}(6)
  (2015).
\newblock \doi{10.1103/physrevd.92.064011}.
\newblock \urlprefix\url{http://dx.doi.org/10.1103/PhysRevD.92.064011}

\bibitem{lisa_gaps}
Q.~Baghi, J.I. Thorpe, J.~Slutsky, et~al., Phys. Rev. D \textbf{100}, 022003
  (2019).
\newblock \doi{10.1103/PhysRevD.100.022003}.
\newblock \urlprefix\url{https://link.aps.org/doi/10.1103/PhysRevD.100.022003}

\bibitem{1975STIN...7714318B}
T.E. {Barnard}.
\newblock {The maximum entropy spectrum and the Burg technique. Technical
  report no. 1: Advanced signal processing}.
\newblock NASA STI/Recon Technical Report N (1975)

\bibitem{Vos}
K.~Vos.
\newblock {A Fast Implementation of Burg's Algorithm}.
\newblock \url{https://opus-codec.org/docs/vos\_fastburg.pdf} (2013)

\end{thebibliography}

\begin{appendix}
\section{Details of PSD computation} \label{sec:appendix}
\subsection{MESA solution}\label{sec:MESA_solution}
We derive the expression for the MAXENT spectral estimator following the approach proposed by \cite{burg1975maximum}.
Unlike the standard approach, we do not enforce the constraints in Eq.~\eqref{eq:MaxConstraint} with the standard Lagrange Multipliers approach.
We write instead the PSD $S(f)$ as the Fourier Transform of the sample autocorrelation function: 
\begin{equation}
    S(f) = \frac{1}{2 Ny}\sum_{n = -\infty}^{\infty} \bar r_n e^{- \imath 2 \pi n \Delta t},
\end{equation}
and, plugging it in the entropy gain expression eq.~\eqref{eq:EntropyGain}, we obtain:
\begin{equation}
    \Delta H = \int_{-Ny}^{Ny}  
    \log\left(\frac{1}{2 Ny}\sum_{n = -\infty}^{\infty} \bar r_n e^{-\imath 2 \pi f n \Delta t} 
    \right) df.
\end{equation}
Note that this expression already takes into account the constraints in eq.~\eqref{eq:MaxConstraint}.

We now introduce a set of coefficients $\lambda_s$, defined as the derivative of $\Delta H$ with respect to the autocorrelation function $r_s$.
Explicitly they are:
\begin{equation} \label{eq:lamdas}
      \lambda_s \coloneqq \frac{\delta H}{\delta \bar r_s} = \frac{1}{2Ny}\int_{-Ny}^{Ny} S(f)^{-1}e^{-\imath 2 \pi f s \Delta t } df
\end{equation} 
and we will show that $S(f)^{-1}$ can be written as a Fourier Expansion in terms of such coefficients. Then, the determination of the values for the $\lambda_s$ uniquely solves the problem of power-spectral density estimation.

Some properties for the coefficients can be worked out easily. First, since $S(f)$ is real, the $\lambda_s$ show the property 
\begin{equation}
	\nonumber 
	\lambda_s = \lambda_{-s}^*. 
\end{equation}
The second property is obtained considering that the autocorrelation function $r_n$ can only be computed for a finite time interval $n \in [-N, N]$ and that the PSD estimation must not depend on the unavailable values $r_n$: this is part of the constraint in eq.~\eqref{eq:MaxConstraint}
This requirement can be implemented as:  
\begin{equation}\nonumber 
    \frac{\delta H}{\delta \bar r_s} = 0 \text{ for } \vert s \vert > N,
\end{equation}
that means 
\begin{equation}
\nonumber 
\lambda_s = 0 \text{ for } \vert s \vert > N. 
\end{equation}

From Eq.~\eqref{eq:lamdas} and from the properties above, is easily seen from the properties of the Fourier transform that $S(f)$ can be expressed via a Fourier Series 
\begin{equation}\label{eq:PSDconstraint}
    S(f)^{-1} = \sum_{s = -N}^N \lambda_s e^{-\imath 2 \pi f s \Delta t}.
\end{equation}
Defining $z = e^{-\imath 2 \pi f \Delta t}$ the previous Fourier expansion becomes a Laurent Polynomial in $z$: 
\begin{equation}
    \label{eq:zExp}
    S(f)^{-1} = \lambda_0 + \sum_{s = 1}^N \lambda_s z^s + \sum_{s = 1}^N \lambda^*_s z^{-s}.
\end{equation}
It is easy to show that if $z_0$ is a root for the polynomial $(z_0^*)^{-1}$ is also a root: for every root laying outside the unit circle there will be another root inside of it and vice-versa. These properties allow us to rewrite the Fourier expansion \eqref{eq:zExp} as \cite{1975STIN...7714318B}:
\begin{equation}\label{eq:MESApsd_appendix}
    S(f) = \frac{P_N \Delta t}{\left(\sum_{s=0}^N a_s z^z\right)\left(\sum_{s = 0}^N a^*_s z^{-s}\right)}
\end{equation}
with $a_0 = 1$ and $\Delta t$ the uniform sampling interval for the time series. The vector obtained as $(1, a_1, \dots, a_N)$ is the prediction error filter. The power spectral density $S(f)$ is uniquely determined if both the prediction error filter and $P_N$ coefficients are computed.

To compute the $a_s$ is convenient to plug into Eq.~\eqref{eq:MaxConstraint} the Laurent Polynomial exansion for $S(f)$ eq.~\eqref{eq:MESApsd_appendix} and then integrating over $z$ (taking values on ${\mathbb S^1}$). In this way the equation becomes:
\begin{equation}
    \label{eq:contourintegral}
   \frac{P_N}{2 \pi \imath} \oint _{\mathbb S^1}\frac{z^{-s - 1}}{\sum_{n = 0}^N a_n z^n \sum_{n = 0}^N a^*_n z^{-n}}dz = \bar r_s. 
\end{equation}
Substituting $s \to s - r$, multiplying by $a^*_s$ and summing over $s$, the previous equation becomes 
\begin{equation}
    \label{eq:errorFilter}
    \sum_{s = 0}^N a_s \bar r_{s - r} = \frac{P_N}{2 \pi \imath}\oint \frac{z^{r -1}}{\sum_{s = 0}^N a_s z^s}dz
\end{equation}
For a wide-sense stationary processes, all the poles lay outside the unit circle so that the previous integral can be easily computed obtaining the following, well known, equations: 
\begin{align}
    \label{eq:errorFilter1}
    \sum_{s = 0}^N a_s \bar r_{r - s} &= P_N \quad \text{ if } r = 0 \\ \label{eq:errorFilter2}
    \sum_{s = 0}^N a_s \bar r_{r - s} & = 0 \qquad \text{ if } r \neq 0.
\end{align}

\subsection{Levinson recursion} \label{sec:LevinsonRecursion}
The solution of the Eqs.~(\ref{eq:errorFilter1}-\ref{eq:errorFilter2}) fully determines the functional form of the power spectral density estimator \eqref{eq:MESApsd_appendix}.
The method for solving the equations is called the Levinson-Durbin recursion \cite{doi:10.1002/sapm1946251261} and it is described in the following.
For each order $N$ of the iteration we define the quantities:
\begin{align}
\Delta_N &= \sum_{n = 0}^{N} a_n \bar{r}_{N - n + 1} \\ 
c_N &= - \frac{\Delta_N}{P_N},
\end{align}

The Levinson recursion computes the $N$th order quantities given the $N-1$th order quantities: 
\begin{equation} \label{eq:Levinson1}
P_N = P_{N -1}\left(1 - \vert c_{N - 1} \vert ^2\right)
\end{equation}
and 
\begin{equation} \label{eq:Levinson2}
\begin{pmatrix}
1 \\ a_1 \\ \vdots \\ a_{N - 1} \\ a_N
\end{pmatrix}
= 
\begin{pmatrix}
1 \\ b_1 \\ \vdots \\ b_{N -1} \\ 0
\end{pmatrix}
+ c_{N-1}
\begin{pmatrix}
0 \\ b_{N -1}^* \\ \vdots \\ b^*_1 \\ 1
\end{pmatrix}. 
\end{equation}
where $b$ holds the value of the $a_s$ coefficients at order $N-1$. 
The 0-th order element can be easily initialized reminding that $a_0 = 1$ (always) and that $P_0$ can be determined from \eqref{eq:errorFilter1}.
Its values turns out to be: 
\begin{equation}
P_0 = R(0),
\end{equation}
$\Delta_0$ and $c_0$ are uniquely determined from their definitions and they are:
\begin{equation}
\Delta_0 = R(1); \quad c_0 = -\frac{R(1)}{R(0)}. 
\end{equation}

These expressions allow us to compute $\vec a$ and $P_N$ to any order by simply iterating \eqref{eq:Levinson1} and \eqref{eq:Levinson2}. Substituting them in equation \eqref{eq:MESApsd_appendix} the problem of the estimation for the power spectral density via maximum entropy principle is solved.
Burg's method for spectral analysis is solved via Levinson is implemented in the released \texttt{memspectrum} package.
Another faster recursion method is available in \cite{Vos} and it is also available in \texttt{memspectrum}.

\section{Computational Time for both MESA methods and Welch}\label{sec:computationaltimes}
In this appendix we shortly introduce the computational times required by the MESA method (considering both the standard and Fast implementation) and the Welch method. They are just inserted to give an idea of what the time differences between the methods are. These are obtained via the python \%timeit special function, run on a personal machine 
\begin{table}[H]
\begin{tabular}{ |p{1.5cm}||p{2cm}|p{2cm}|p{2cm}|  }
 \hline
 \multicolumn{4}{|c|}{Computational times} \\
 \hline
 Batch Length& MESA std &MESA Fast&Welch\\

 \hline
 1s   &  22  ± 1.22 ms   &$19.6$ ± $0.62$ ms &  335 ± 9.24 µs\\
 5s &   158  ± 21.7 ms   & 42.4 ± 0.35 ms   &839 ± 4.61 µs \\
 10s & 187  ± 11.6 ms & 51.5 ± 3.67 ms&  1.74  ± 0.06 µs \\
 100s    & 1.96 ± 0.34 ms  &205  ± 5.09 &  18.8 ± 0.14 ms\\
 1000s&  17.1  ± 0.61 ms  &1.33  ± 0.02 ms &235 ± 3.69 ms\\

 \hline
\end{tabular}
\caption{Comparison of the computational times for the estimate of the power spectral densities with our implementation of MESA (both standard and Fast implementations) and Welch's method}
\label{tb1:ComputationalTimes}
\end{table}

\section{Full posterior distribution for GW150914} \label{sec:appendix_b}
\begin{figure*}
	\caption{Full posterior distribution for GW150914 when marginalising over the AR process orders.}
	\label{fig:gw150914_full_pe}
	\includegraphics[width=\textwidth,keepaspectratio]{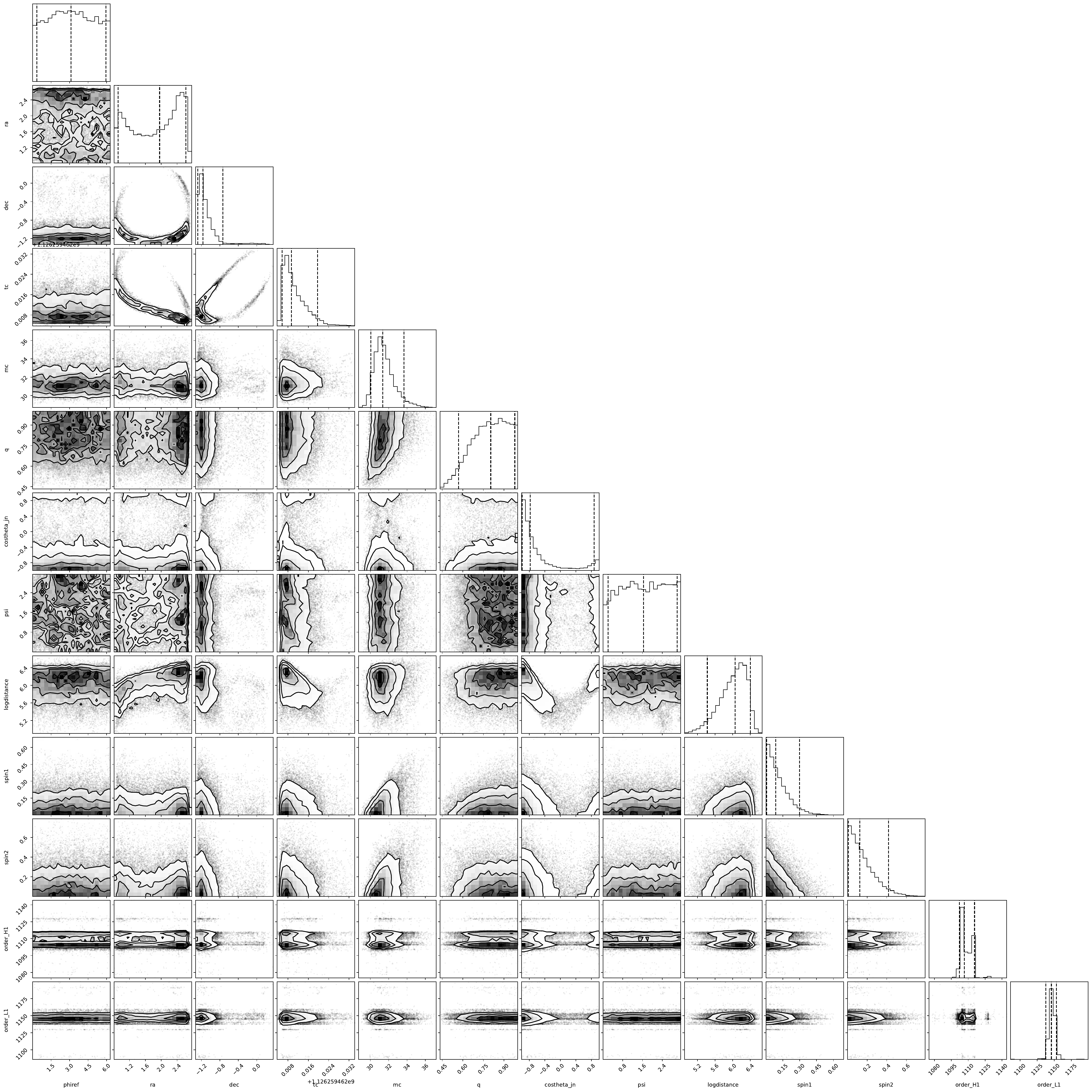}
\end{figure*}
\end{appendix}

\end{document}